# Quantum percolation based dynamic propagation connectivity for critical-area identification in transport networks

Junxiang Xu[1,*], Chence Niu[2], Vinayak Dixit[1], Divya Jayakumar Nair[1],Tingting Zhang[3,4]

*1. Research Centre for Integrated Transport Innovation (rCITI), School of Civil and Environmental Engineering, The University of New South Wales, Kensington, UNSW Sydney, NSW, 2052, Australia*

*2. Guangdong Basic Research Center of Excellence for Ecological Security and Green Development, Key Laboratory for City Cluster Environmental Safety and Green Development of the Ministry of Education, School of Ecology, Environment and Resources, Guangdong University of Technology, Guangzhou, 510006, China*

*3. Key Laboratory for Vehicle Emission Control and Simulation of Ministry of Ecology and Environment, Chinese Research Academy of Environmental Sciences, Beijing, China*

*4. Vehicle Emission Control Center, Chinese Research Academy of Environmental Sciences, Beijing, China*

**corresponding author*:** junxiang.xu@unsw.edu.au

**Abstract:** Transport networks often lose functionality through gradual degradation in link operating conditions before topological disconnection occurs. Link-centred and binary percolation measures identify important facilities or connectivity failures, but they provide limited information on which spatial areas cause the largest loss of network-wide propagation capability. This paper develops a Dynamic Propagation Connectivity (DPC) metric based on quantum percolation for critical-area identification in transport networks. Time-varying link travel times are converted into continuous propagation strengths, which define a Hermitian propagation operator at each observation time. Candidate regions are then evaluated by a regional degradation experiment that measures the resulting loss of DPC. The method is applied to a benchmark Sioux Falls network and six Florida road networks during the post-Hurricane Irma disruption and recovery period, using 1,281 five-minute observation times. The benchmark confirms that the regional DPC score identifies a predefined structurally critical corridor. In the Florida networks, the identified critical areas differ from regions selected by link count, local degradation, edge betweenness, algebraic connectivity, and classical percolation. In Networks 1 to 4, DPC and classical percolation rankings have negative Spearman correlations, showing that continuous propagation degradation and binary fragmentation reveal different vulnerability patterns. Robustness tests under alternative travel time scaling, degradation strength, and grid size show stable results, with mean rank agreement between 0.84 and 0.96. The findings extend transport resilience analysis based on percolation from binary connectivity loss to continuous propagation degradation and provide a spatial diagnostic tool for regional monitoring, emergency planning, and recovery prioritisation.

**Keywords:** Quantum percolation, dynamic propagation connectivity, critical-area identification, transport network resilience, time-varying travel time, network vulnerability

## 1. Introduction

### *1.1 Research background and motivation*

Natural hazards often disrupt transport networks over spatially contiguous areas (Mattsson and Jenelius, 2015). Hurricane wind fields and rainfall degrade corridors along the storm path. Floods inundate multiple roads across low-lying areas. Earthquakes damage bridges and road segments along fault lines simultaneously. In these scenarios, multiple links within a geographically concentrated area experience functional degradation at the same time. Transport agencies also organise hazard response decisions around spatial units, such as evacuation zones, traffic control areas, administrative districts, and repair corridors (Jenelius and Mattsson, 2012). Identifying the spatial areas whose degradation causes the largest reduction in overall network performance directly supports spatial prioritisation and resource deployment during hazard events.

Most transport network vulnerability studies use individual links or nodes as the analytical unit and produce importance rankings of single links or link sets (Jenelius et al., 2006; Sullivan et al., 2010; Bell et al., 2017). These rankings provide valuable diagnostic information for local bottlenecks. Hazard disruptions often affect spatially contiguous areas, and transport agencies organise response actions around spatial units. The analytical unit in existing studies therefore differs from the decision unit in practice. Converting link rankings into spatial action plans requires additional aggregation and manual interpretation. Critical-area identification closes this gap by aligning the analytical unit directly with the decision unit and answering an operational question: which spatial areas, once degraded, cause the largest reduction in overall network performance. This alignment produces actionable spatial rankings that transport agencies can use directly for zonal management, repair scheduling, and evacuation monitoring during hazard response.

### *1.2 Current research*

#### *1.2.1 Critical link identification*

Critical link identification has developed along two main lines. The first line evaluates the vulnerability and importance of individual links or nodes. Bell (2000) proposed a game theory approach to measure transport network performance reliability, linking network performance to the worst-case link degradation state. Jenelius et al. (2006) defined link importance and exposure indicators to assess the effect of single link closures on regional accessibility. Sullivan et al. (2010) developed a network robustness index to identify critical road segments and measure system ability to withstand disruptions. Bell et al. (2017) applied capacity-weighted spectral analysis to vulnerability assessment and identified the links with the largest influence on network connectivity through changes in algebraic connectivity. The second line formulates critical link set identification as a combinatorial optimisation problem, using network interdiction models, bi-level programming, and mixed-integer programming. The objective function typically maximises the increase in total system travel time or system loss under user equilibrium conditions. Faturechi and Miller-Hooks (2014) proposed a stochastic optimisation framework to quantify roadway travel time resilience under disaster conditions, measuring how disruption scenarios and recovery actions jointly affect system performance. Recent work from network science has further enriched the analytical tools for critical link identification. Ganin et al. (2017) analysed the trade-off between resilience and efficiency in transport networks and identified network components with critical influence on overall system performance. Saberi et al. (2020) found that traffic congestion spreads through urban road networks as a simple contagion process and identified the critical bottlenecks that trigger congestion cascades. These two main lines and recent interdisciplinary contributions have established a mature methodological foundation spanning single link evaluation to link set optimisation.

#### *1.2.2 Percolation and network connectivity*

Percolation theory analyses network vulnerability from a connectivity perspective. Classical percolation theory originated from the mathematical description of fluid propagation through random media by Broadbent and Hammersley (1957). Albert et al. (2000) and Cohen et al. (2000) introduced percolation analysis to complex networks and studied the collapse of network connectivity under random failures and targeted attacks. In transport networks, Li et al. (2015) discovered percolation phase transitions in dynamical traffic networks, where the evolution of congestion bottlenecks drives the network from a connected state to a fragmented state. Zeng et al. (2019) further revealed the switching mechanism between two critical percolation modes in urban traffic dynamics. Wang et al. (2019) showed that local floods can induce large-scale abrupt failures of road networks through a percolation mechanism. Hamedmoghadam et al. (2021) applied percolation analysis of heterogeneous flows to infrastructure networks and identified the bottleneck structures that constrain overall network connectivity. Zhu et al. (2025) applied percolation to evaluate urban rail transit resilience under different operation schemes, extending percolation methods to multimodal transport systems.

In transport percolation analysis, each link is classified as functional or failed according to a speed threshold. The binary state of link $(i, j)$ at observation time $\tau$ is defined as:

$$b_{ij}(\tau) = \begin{cases} 1, & v_{ij}(\tau) \geq v^* \\ 0, & v_{ij}(\tau) < v^* \end{cases} \tag{1}$$

where $v_{ij}(\tau)$ is the observed speed and $v^*$ is the speed threshold (Li et al., 2015; Zeng et al., 2019). This threshold retains links with observed speed at or above $v^*$ and removes the rest. The connectivity of the resulting binary network is measured by the relative size of the largest connected component:

$$S_{\max}(\tau) = \frac{|C_{\max}(G'(\tau))|}{N} \tag{2}$$

where $G'(\tau)$ is the thresholded binary network, $C_{\max}(G'(\tau))$ is its largest connected component, and $N$ is the total number of nodes. A value of $S_{\max}(\tau)$ close to 1 indicates that the network maintains overall connectivity. A sharp decline indicates network fragmentation.

Percolation-related spectral methods have also been used to measure network connectivity. For example, researchers introduced algebraic connectivity as a spectral measure of graph connectivity and applied capacity-weighted algebraic connectivity to transport network vulnerability assessment (Fiedler, 1973; Bell et al., 2017). Classical percolation and algebraic connectivity both operate under a binary failure assumption. At any given threshold, each link is either fully retained or fully removed. The threshold can be varied to explore different percolation states, but the link state remains binary at each threshold value.

*1.2.3 Critical area identification*

Compared with the extensive literature on critical link identification, studies that use spatial areas as the analytical unit remain limited. Taylor et al. (2006) proposed accessibility methods for vulnerability analysis of strategic road networks and linked vulnerability to spatial accessibility loss. Sohn (2006) evaluated the significance of highway links under flood damage conditions and incorporated the spatial distribution of hazard impacts into vulnerability assessment. Jenelius (2009) analysed the geographical disparities of road network vulnerability from network structure and travel pattern perspectives and revealed the spatially uneven distribution of vulnerability across regions. Jenelius and Mattsson (2012) proposed the most direct framework for area-covering disruption analysis. Their method partitions the study area into regular grid cells, removes all links inside each grid cell, and ranks candidate areas by the resulting increase in total system travel time. This grid enumeration approach provides a systematic spatial framework for critical area identification.

These existing studies share two common features. First, links inside a candidate area are fully removed from the network during the disruption test. Second, the evaluation metric is typically total system travel time increase or largest connected component loss. Under natural hazard conditions, link functionality degrades continuously and progressively. Link speeds decline to varying degrees, and different links within the same area experience different levels of degradation. The full removal assumption cannot distinguish these continuous degradation states. Furthermore, enumeration of candidate areas requires repeated link removal and system performance recalculation for each candidate region, and the computational cost grows with network size and the number of candidate areas.

Figure 1 illustrates the conceptual difference between critical link identification and critical area identification. Critical link identification uses individual links as the analytical unit and evaluates the effect of single link failure or link set removal on system performance. Critical area identification uses spatial areas as the analytical unit and evaluates the effect of regional link degradation on overall network performance. The output of critical area identification corresponds directly to the spatial decision units used in transport management.

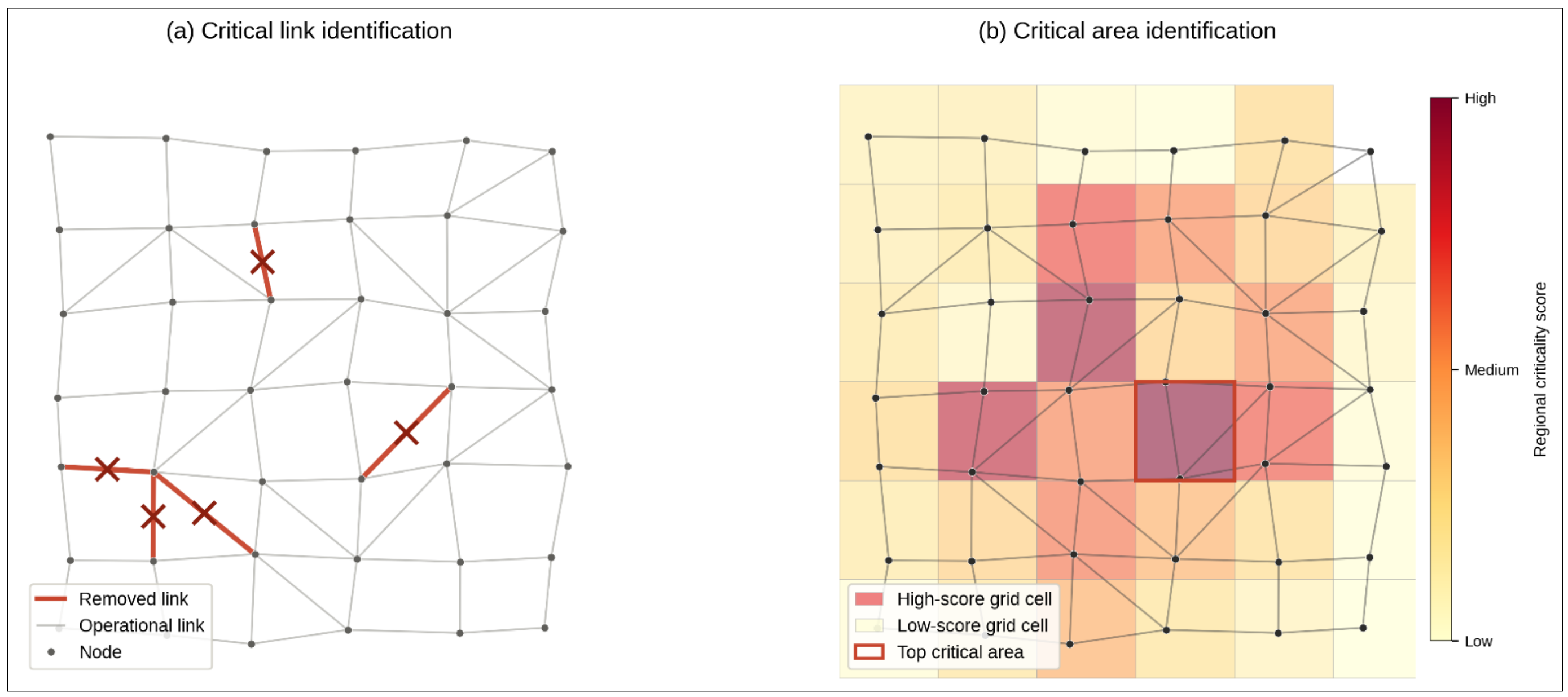

Figure 1. Conceptual comparison of critical link identification and critical area identification.

*1.3 Research gap*

The literature reviewed above reveals three interrelated methodological gaps. First, existing vulnerability analysis methods predominantly use individual links or link sets as the analytical unit. Critical area identification directly aligns the analytical unit with the spatial decision units used in hazard response but remains underexplored. The few available methods rely on exhaustive enumeration of candidate areas with full link removal inside each area. Second, classical percolation and existing area vulnerability methods both adopt a binary failure model and evaluate static disruption scenarios. Each link is either fully retained or fully removed, and each candidate area is tested once under a fixed removal condition. Natural hazards produce continuous and progressive link degradation that evolves over time, and current area vulnerability methods do not capture this temporal dimension. Third, existing area vulnerability metrics such as total system travel time increase require origin-destination demand matrices and traffic assignment models. During natural hazard events, demand data are often unavailable or unreliable. Area vulnerability analysis still lacks a propagation connectivity measure that accommodates continuous functional degradation and operates directly on observed link travel time data without requiring demand inputs.

These three gaps point to a common methodological need. A new framework should use spatial areas as the analytical unit, accommodate continuous and time-varying functional degradation, and provide a demand-free propagation coverage measure beyond binary connectivity.

*1.4 Quantum percolation and propagation connectivity*

Quantum percolation theory provides a theoretical foundation for addressing the gaps identified above. Classical percolation evaluates connectivity through binary link states and topological connectedness. Quantum percolation evaluates connectivity through the propagation ability of wave functions across disordered networks. This shift from binary connectivity to continuous propagation coverage offers a natural framework for analysing transport network vulnerability under continuous functional degradation.

Quantum percolation originated in condensed matter physics. Anderson (1958) established the theory of localisation in disordered systems and proved that wave functions become spatially confined when structural disorder reaches a sufficient level. Mülken and Blumen (2011) provided a systematic review of continuous-time quantum walk models on complex networks and identified localisation as a central feature of propagation in disordered structures. Biamonte et al. (2019) reviewed the theoretical development from classical networks to quantum networks and highlighted the decisive role of spectral structure in network propagation. Xu et al. (2021) experimentally verified quantum transport behaviour on fractal networks and demonstrated the influence of network topology on propagation properties.

Within the quantum percolation framework, a network may remain topologically connected while its propagation states become localised due to structural degradation. The spectral structure of the propagation operator determines whether propagation states extend across the network or concentrate on a small subset of nodes. This continuous propagation degradation mechanism differs fundamentally from the binary connectivity threshold in classical percolation. To intuitively compare the attributes of the two types of percolation theory, Table 1 compares the features of classical percolation theory and quantum percolation theory across multiple analytical dimensions.

Table 1. Comparison between classical percolation and quantum percolation.

| Feature | Classical percolation | Quantum percolation |
|---|---|---|
| Binary connectivity | √ | |
| Connectivity determined by wave-function propagation | | √ |
| Giant connected component as a structural order parameter | √ | √ |
| Random occupation probability as structural basis | √ | √ |
| Explicit propagation mechanism | | √ |
| No explicit time evolution (static configuration ensemble) | √ | |
| Localisation transition | | √ |
| Dominated by geometric connectivity | √ | |
| Dominated by spectral structure (eigenvalues/eigenstates) | | √ |
| Captures "connected but slow / weakly propagating" states | | √ |
| Capable of indicating propagation paths and effective spreading range | | √ |

*1.5 Research contribution*

This paper makes four contributions to transport network vulnerability analysis.

(1) This paper proposes a regional Dynamic Propagation Connectivity (DPC) score for critical area identification. The method extends global DPC to the spatial area level by applying continuous functional degradation to candidate regions and measuring the resulting DPC loss.

(2) This paper establishes a data-driven conversion from observed link travel times to continuous propagation strengths. This conversion enables the propagation framework to operate directly on observed traffic data without requiring origin-destination demand matrices or traffic assignment models.

(3) This paper achieves time-varying critical area tracking. The regional DPC score is computed independently at each observation time, producing a temporal sequence of regional criticality. Three summary indicators capture the persistence, peak impact, and ranking stability of each candidate area over the observation period.

(4) This paper validates the proposed framework through a benchmark experiment on the Sioux Falls network and a large-scale empirical application on six regional road networks in Florida during Hurricane Irma, covering network sizes from 154 to 7,827 directed links and 1,281 observation times. The empirical analysis systematically compares the regional DPC score with five baseline indicators and demonstrates that regional DPC identifies different critical areas from classical percolation, algebraic connectivity, link count, local degradation, and edge betweenness.

The remainder of this paper is organised as follows. Section 2 presents the methodological framework, including global DPC, link-travel-time-driven propagation strength, regional degradation, and the regional DPC score. Section 3 describes the six Florida regional traffic network data sets and the experimental settings. Section 4 reports the empirical results, covering benchmark validation on the Sioux Falls network, global DPC evolution, critical area identification, comparison with baseline indicators, and robustness analysis. Section 5 discusses the findings and their implications. Section 6 concludes the paper.

## 2. Methodological framework

### *2.1 Global DPC*

Consider a transport network observed at time $\tau$. The network state is represented by $G(\tau)=\left(V,E,A(\tau)\right)$, where $V=\{1,2,...,N\}$ is the node set, $E\subseteq V\times V$ is the directed link set, and $A(\tau)=\left[a_{ij}(\tau)\right]$ is the effective propagation strength matrix. The value $a_{ij}(\tau)\geq 0$ describes the ability of link $(i,j)$ to support propagation at time $\tau$. In this paper, $\tau$ denotes the observation time in the real traffic data, it corresponds to a five-minute link travel time observation.

The observed link travel times provide a time-varying traffic information stream. This information stream is converted into the effective propagation strength $a_{ij}(\tau)$. The conversion from link travel time to propagation strength is introduced in Section 2.2. This subsection first defines propagation connectivity for a general network state $G(\tau)$.

For a fixed network state $G(\tau)$, each node $i\in V$ is associated with a basis state $|i\rangle$, with $|i\rangle$ denoting its dual vector. The propagation operator is defined as:

$$H(\tau)=\sum_{(i,j)\in E}a_{ij}(\tau)|i\rangle\langle j| \tag{3}$$

Under the column-vector convention, $|i\rangle\langle j|$ maps the component at node $j$ to node $i$, so $a_{ij}(\tau)$ is the $(i,j)$ coupling entry of the propagation operator. This operator embeds the functional state of the transport network into the coupling structure of the propagation process. This construction follows the use of Hamiltonian-type coupling matrices in continuous-time quantum walk models for transport on complex networks (Mülken and Blumen, 2011).

Transport networks are usually directed. As a result, $H(\tau)$ is not necessarily Hermitian. To support spectral propagation analysis, this paper uses the Hermitian symmetrised operator:

$$\bar{H}(\tau)=\frac{H(\tau)+H(\tau)^{\dagger}}{2} \tag{4}$$

where $(\cdot)^{\dagger}$ denotes the conjugate transpose. After this symmetrisation, $H(\tau)$ is used to denote $\bar{H}(\tau)$ for notational simplicity. Hermitian matrix representations are commonly used to obtain real spectra for directed and mixed graphs (Guo and Mohar, 2017).

Two-time scales are used in the analysis. The observation time $\tau$ indexes the traffic information stream. The propagation time $t$ describes the internal evolution of a propagation state under a fixed network state $G(\tau)$. Once $\tau$ is fixed, the network state is treated as constant during the propagation process. This gives a quasi-static representation of each observed traffic state:

$$|P(G,0;\tau)\rangle=\frac{1}{\sqrt{N}}\sum_{i\in V}|i\rangle \tag{5}$$

This choice assigns the same initial amplitude to all nodes and avoids imposing a preferred propagation origin.

Under the fixed network state $G(\tau)$, let $|P(G,t;\tau)\rangle$ denote the propagation state at propagation time $t$. Its evolution follows:

$$\mathrm{i}\frac{d}{dt}|P(G,t;\tau)\rangle=H(\tau)|P(G,t;\tau)\rangle \tag{6}$$

The corresponding solution is:

$$|P(G,t;\tau)\rangle=\exp[-\mathrm{i}H(\tau)t]|P(G,0;\tau)\rangle \tag{7}$$

The spatial extent of the propagation state is measured by the participation index:

$$\Pi(\tau,t)=\frac{\left(\sum_{i\in V}|P_i(G,t;\tau)|^2\right)^2}{\sum_{i\in V}|P_i(G,t;\tau)|^4} \tag{8}$$

where $P_i(G,t;\tau)$ is the component of the propagation state at node $i$. A large value of $\Pi(\tau,t)$ means that the propagation state covers many nodes. A small value means that the propagation state is concentrated on a small part of the network. Participation measures are widely used to distinguish spatially extended states from localised states in spectral and quantum transport analysis (Evers and Mirlin, 2008). In the present transport network setting, $\Pi(\tau,t)$ gives the effective number of nodes reached by the propagation state at propagation time $t$.

The global DPC at observation time $\tau$ is defined as the time-averaged participation index:

$$DPC(\tau)=\lim_{L\to\infty}\frac{1}{L}\int_0^L\Pi(\tau,t)dt \tag{9}$$

where $L$ is the averaging horizon over the internal propagation time. This measure gives the effective number of nodes covered by the propagation process under the traffic network state observed at $\tau$. The term dynamic refers to the variation of $DPC(\tau)$ across observation times. For each fixed $\tau$, the average over the internal propagation time $t$ extracts the long-run propagation coverage associated with that observed network state. Note that Eq. (9) is taken over the propagation time $t$, not over the observation time $\tau$. For a given traffic observation $\tau$, the network state $G(\tau)$ is fixed. The propagation state may still oscillate with $t$ under this fixed network state. The long-time average removes this dependence on a particular propagation instant and returns a stable measure of propagation coverage for $G(\tau)$.

The definitions in this subsection provide the global propagation measure used in the rest of the paper. The following subsections specify how link travel time information determines $a_{ij}(\tau)$, and how the resulting global measure can be extended to identify spatial regions that are critical to network-wide propagation connectivity.

*2.2 Link-travel-time-driven propagation strength*

This subsection specifies how observed link travel time information is converted into the effective propagation strength $a_{ij}(\tau)$. For each directed link $(i,j)\in E$, let $T_{ij}(\tau)$ denote the observed travel time of the link at observation time $\tau$. A shorter travel time indicates that the link can support propagation more efficiently under the observed traffic state. A longer travel time indicates a weaker propagation condition caused by congestion, hazard impacts, or other operational disruptions.

Observed travel times are converted into dimensionless propagation strengths because raw travel times are affected by link length, road class, and free-flow conditions. The same travel time value may represent different levels of functionality on different links. To control for these link-specific differences, this paper defines a scaling travel time $T_{ij}^{0}$ for each link. Let $\mathrm{T}_{ij}^{+}$ denote the set of valid observation times for link $(i,j)$. The scaling travel time is defined as:

$$T_{ij}^{0}=Q_{q}\left(\left\{T_{ij}(\tau):\tau\in \mathrm{T}_{ij}^{+}\right\}\right) \tag{10}$$

where $Q_{q}(\cdot)$ is the empirical $q$-quantile. The main analysis uses $q=0.05$, so $T_{ij}^{0}$ is estimated as the fifth percentile of the observed travel times on link $(i,j)$. This quantity serves as a link-specific scaling factor for normalisation. In an operational implementation, the same scaling factor can be calibrated from historical link travel time records.

The effective propagation strength of link $(i,j)$ at observation time $\tau$ is then defined as:

$$a_{ij}(\tau)=\min\left\{1,\frac{T_{ij}^{0}}{T_{ij}(\tau)}\right\} \tag{11}$$

Equation (11) maps link travel time information into a dimensionless propagation strength. When the observed travel time is close to the scaling travel time, $a_{ij}(\tau)$ is close to 1 and the link retains a strong propagation condition. When the observed travel time increases, $a_{ij}(\tau)$ decreases and the link contributes less to propagation. The upper bound of 1 prevents unusually small travel time observations from producing propagation strengths above the fully available state.

The resulting matrix $A(\tau)=\left[a_{ij}(\tau)\right]$ represents the time-varying traffic-state information contained in the link travel time data. Once $A(\tau)$ is obtained, each observation time $\tau$ gives a propagation operator $H(\tau)$ and a global propagation connectivity value $DPC(\tau)$. The next subsection extends this construction from global monitoring to spatial critical-area identification.

*2.3 Spatial regions and regional degradation*

This subsection defines the spatial objects used for critical-area identification. The study area is represented by $M$ candidate regions:

$$\mathrm{R}=\left\{R_{1},R_{2},\ldots,R_{M}\right\} \tag{12}$$

Each region $R_{m}, m=1,2,\ldots,M$ is a spatial unit with a defined boundary. The empirical analysis uses regular spatial grids, which provide transparent and reproducible candidate regions. Other spatial divisions, such as traffic management zones or administrative areas, can be used when such boundaries are available.

To link the spatial partition with the transport network, let $z_{ij}$ denote the representative geographic coordinate of link $(i,j)$.

In the empirical data, $z_{ij}$ is taken from the link-level latitude and longitude fields, which provide one fixed spatial reference point for each link across observation times. The set of links assigned to region $R_m$ is:

$$E_{R_m} = \{(i,j) \in E : z_{ij} \in R_m\} \tag{13}$$

Equation (13) assigns each link to a candidate region according to its geographic coordinate. The resulting set $E_{R_m}$ contains all links whose coordinates fall within the boundary of $R_m$.

**Definition 1 (Regional degradation).** For a candidate region $R_m \in \mathrm{R}$ and a degradation level $\alpha \in [0,1]$, the regional degraded network state at observation time $\tau$ is denoted by:

$$G^{(-R_m,\alpha)}(\tau) = \left(V, E, A^{(-R_m,\alpha)}(\tau)\right) \tag{14}$$

where $A^{(-R_m,\alpha)}(\tau) = [a_{ij}^{(-R_m,\alpha)}(\tau)]$, and

$$a_{ij}^{(-R_m,\alpha)}(\tau) = \begin{cases} (1-\alpha)a_{ij}(\tau), & (i,j) \in E_{R_m} \\ a_{ij}(\tau), & (i,j) \notin E_{R_m} \end{cases} \tag{15}$$

The parameter $\alpha$ controls the diagnostic degradation level applied to links in $R_m$. When $\alpha = 0$, the regional degraded state coincides with the observed network state. When $\alpha = 1$ the propagation strengths of links in $R_m$ are fully removed. Intermediate values represent partial degradation of the links assigned to the region.

The raw regional degraded propagation operator is:

$$\hat{H}^{(-R_m,\alpha)}(\tau) = \sum_{(i,j)\in E} a_{ij}^{(-R_m,\alpha)}(\tau)|i\rangle\langle j| \tag{16}$$

The corresponding Hermitian regional degraded propagation operator is:

$$H^{(-R_m,\alpha)}(\tau) = \frac{\hat{H}^{(-R_m,\alpha)}(\tau) + \hat{H}^{(-R_m,\alpha)}(\tau)^{\dagger}}{2} \tag{17}$$

This operator represents the propagation structure obtained after applying the regional degradation operation to $R_m$ at observation time $\tau$. The next subsection compares the DPC of the observed network with the DPC obtained after degrading $R_m$, and then defines the regional DPC loss.

### *2.4 Regional DPC impact and critical-area score*

The criticality of a region is measured by the DPC response caused by degrading that region. For a fixed observation time $\tau$, the observed network has an original DPC value $DPC(\tau)$. For a candidate region $R_m$ and degradation level $\alpha$, the degraded operator $H^{(-R_m,\alpha)}(\tau)$ defined in Eq. (17) gives the corresponding degraded DPC value.

**Definition 2 (Regional DPC impact and critical-area score).** Let $\mathrm{D}(\cdot)$ denote the DPC functional induced by Eqs. (6) to (9) for the fixed node set $V$ and the initial state in Eq. (5). The observed and regionally degraded DPC values are:

$$\mathrm{D}\left(H(\tau)\right) = DPC(\tau),\ \mathrm{D}\left(H^{(-R_m,\alpha)}(\tau)\right) = DPC^{(-R_m,\alpha)}(\tau) \tag{18}$$

For observation times with $DPC(\tau) > 0$, the signed regional DPC impact is defined as:

$$I_m(\tau,\alpha) = \frac{DPC(\tau) - DPC^{(-R_m,\alpha)}(\tau)}{DPC(\tau)} \tag{19}$$

A positive value of $I_m(\tau,\alpha)$ indicates that degrading $R_m$ reduces global DPC. A value close to 0 indicates a limited global effect. A negative value corresponds to a case where the degraded operator yields a larger DPC value than the observed operator.

The non-negative critical-area score is defined as:

$$S_m(\tau,\alpha) = \max\{0, I_m(\tau,\alpha)\} \tag{20}$$

The score $S_m(\tau,\alpha)$ measures the DPC loss caused by degrading region $R_m$. Larger values indicate regions whose degradation produces stronger losses of global propagation connectivity. Critical areas are obtained by ranking candidate regions in descending order of $S_m(\tau,\alpha)$. In the empirical analysis, the regions with the largest scores are mapped as the identified critical areas.

**Proposition 1 (Finite-difference response to regional degradation).** For a fixed observation time $\tau$ and candidate region $R_m$, define the regional perturbation operator by:

$$\hat{B}_m(\tau) = \sum_{(i,j)\in E_{R_m}} a_{ij}(\tau)|i\rangle\langle j|,\ B_m(\tau) = \frac{\hat{B}_m(\tau)+\hat{B}_m(\tau)^\dagger}{2} \tag{21}$$

Then the regional degraded Hermitian operator satisfies:

$$H^{(-R_m,\alpha)}(\tau) = H(\tau) - \alpha B_m(\tau) \tag{22}$$

Consequently, the signed regional impact is the normalised finite-difference response of DPC to the regional perturbation:

$$I_m(\tau,\alpha) = \frac{\mathrm{D}(H(\tau)) - \mathrm{D}(H(\tau)-\alpha B_m(\tau))}{\mathrm{D}(H(\tau))} \tag{23}$$

If $\phi_m(\alpha;\tau) = \mathrm{D}(H(\tau)-\alpha B_m(\tau))$ is differentiable at $\alpha = 0$, then, as $\alpha \to 0$,

$$I_m(\tau,\alpha) = -\frac{\alpha}{DPC(\tau)} \left.\frac{\partial \phi_m(\alpha;\tau)}{\partial \alpha}\right|_{\alpha=0} + o(\alpha) \tag{24}$$

where $o(\alpha)$ denotes a remainder term satisfying $o(\alpha)/\alpha \to 0$ as $\alpha \to 0$.

**Proof.** Let $\hat{H}(\tau)$ denote the raw observed propagation operator before Hermitian symmetrisation. By Eq. (15), the regional degraded raw operator can be written as:

$$\hat{H}^{(-R_m,\alpha)}(\tau) = \hat{H}(\tau) - \alpha \hat{B}_m(\tau) \tag{25}$$

Applying Hermitian symmetrisation to both sides of Eq. (25) gives:

$$\frac{\hat{H}^{(-R_m,\alpha)}(\tau) + \hat{H}^{(-R_m,\alpha)}(\tau)^\dagger}{2} = \frac{\hat{H}(\tau)+\hat{H}(\tau)^\dagger}{2} - \alpha\frac{\hat{B}_m(\tau)+\hat{B}_m(\tau)^\dagger}{2} \tag{26}$$

Using Eq. (4), Eq. (17), and Eq. (21), Eq. (26) gives Eq. (22). Substituting Eq. (22) into **Definition 2** gives Eq. (23). Finally, the differentiability of $\phi_m(\alpha;\tau)$ at $\alpha = 0$ gives the first-order expansion:

$$\phi_m(\alpha;\tau) = \phi_m(0;\tau) + \alpha \left.\frac{\partial \phi_m(\alpha;\tau)}{\partial \alpha}\right|_{\alpha=0} + o(\alpha) \tag{27}$$

By the definition of $\phi_m(\alpha;\tau)$ and Eq. (18), $\phi_m(0;\tau) = DPC(\tau)$, substituting Eq. (27) into Eq. (19) gives Eq. (24).

**Proposition 1** shows that the proposed regional score is a DPC response to a structured regional perturbation. The score therefore captures the global propagation consequence of degrading a candidate region, rather than the local traffic state within that region alone.

*2.5 Computational workflow*

This subsection summarises the computational workflow used to implement the proposed framework. The workflow starts from a directed transport network, link-level travel time observations, link-level spatial coordinates, a set of candidate regions, a quantile level $q$, and a diagnostic degradation level $\alpha$. Table 2 reports the main steps, operations, and outputs.

Table 2. Computational workflow for critical-area identification.

| Step | Task | Main operation | Output |
|---|---|---|---|
| 1 | Scaling travel time estimation | Estimate $T_{ij}^{0}$ for each link using Eq. (10). | $T_{ij}^{0}$ |
| 2 | Propagation strength construction | Convert $T_{ij}(\tau)$ into $a_{ij}(\tau)$ using Eq. (11). | $A(\tau)$ |
| 3 | Global DPC computation | Construct $H(\tau)$ and compute $DPC(\tau)$ using Eqs. (3) to (9). | $DPC(\tau)$ |
| 4 | Regional link assignment | Assign links to candidate regions using Eq. (13). | $E_{R_m}$ |
| 5 | Regional degradation | Construct $H^{(-R_m,\alpha)}(\tau)$ using Eqs. (15) to (17). | $H^{(-R_m,\alpha)}(\tau)$ |
| 6 | Regional score computation | Compute $DPC^{(-R_m,\alpha)}(\tau), I_m(\tau,\alpha), S_m(\tau,\alpha)$ using Eqs. (18) to (20). | $S_m(\tau,\alpha)$ |
| 7 | Critical-area identification | Rank regions by $S_m(\tau,\alpha)$ and map the highest-scoring regions. | Critical-area map and ranking |

The workflow is repeated for each observation time. This produces a time-varying representation of both global propagation connectivity and regional criticality. The computations for different observation times and candidate regions are separable, which supports parallel implementation for large transport networks.

## 3. Data and experimental design

### *3.1 Study area and empirical data description*

The empirical analysis uses six regional link travel time data sets from Florida during the post-Irma disruption and recovery period. The six networks cover different parts of the Florida road system, including the Space Coast, the Treasure Coast, the south-eastern coastal corridor, Miami-Dade, the Lower Florida Keys, and south-west Florida. This spatial coverage provides a suitable empirical setting for testing whether the proposed quantum percolation-based DPC framework can identify critical areas across transport networks with different sizes, locations, and traffic conditions.

Each data set records the operating state of directed road links at a sequence of observation times. Each record contains a link identifier, origin and destination nodes, link length, a representative geographic coordinate, observation time, link travel time, and average speed. Following the notation in Section 2, the origin and destination nodes define a directed link $(i,j)$, the observed link travel time gives $T_{ij}(\tau)$, the representative geographic coordinate gives $z_{ij}$, and the timestamp gives the observation time $\tau$. The data therefore provide the two inputs required by the proposed method: time-varying link operating conditions and link-level spatial locations. It should be noted that the data provide one fixed geographic coordinate for each link $(i,j)$ rather than full polyline geometry. This coordinate is used only for assigning links to candidate spatial regions in Section 3.3.

The six data sets have the same available observation period, from 11 September 2017 14:35 to 18 September 2017 10:20. The traffic state is recorded at five-minute intervals over the available observation times. The observation timestamps contain temporal gaps, so the analysis treats the 1,281 available timestamps as the discrete observation time set. Within each data set, the link-time panel is complete over these available observation times. Each available timestamp contains all link records in the corresponding regional network.

The six regional networks differ substantially in size. The smallest network contains 154 link records and 125 nodes,

whereas the largest network contains 7,827 link records and 6,081 nodes. Across the six data sets, the empirical sample contains 23,770,236 link-time observations. The invalid record share remains below 0.1 percent in every data set. Invalid records refer to non-positive or non-finite travel time or speed observations. The underlying network in each region is connected. Table 3 summarises the main data characteristics.

Table 3. Summary of the six regional traffic network data sets.

| Network | Region | Directed link records | Nodes | Link-time observations | Available observation times | Mean speed (km/h) | Invalid records (%) | Connectivity (%) |
|---|---|---|---|---|---|---|---|---|
| Network 1 | Palm Bay / Brevard County | 1,606 | 1,339 | 2,057,286 | 1,281 | 54.01 | 0.000 | 100.0 |
| Network 2 | Fort Pierce / Port St. Lucie | 3,253 | 2,681 | 4,167,093 | 1,281 | 54.66 | 0.062 | 100.0 |
| Network 3 | Palm Beach-Fort Lauderdale corridor | 5,407 | 4,522 | 6,926,367 | 1,281 | 44.93 | 0.031 | 100.0 |
| Network 4 | Miami-Dade / Homestead | 7,827 | 6,081 | 10,026,387 | 1,281 | 41.41 | 0.009 | 100.0 |
| Network 5 | Key West / Lower Florida Keys | 154 | 125 | 197,274 | 1,281 | 38.98 | 0.000 | 100.0 |
| Network 6 | Naples / Collier County | 309 | 280 | 395,829 | 1,281 | 61.46 | 0.000 | 100.0 |

Note: Invalid records refer to non-positive or non-finite travel-time or speed observations.

Figure 2 shows the spatial distribution of the six regional traffic networks. Points represent link-level representative geographic coordinates, and colours distinguish the six regional networks used in the empirical analysis. Because the data provide representative link coordinates rather than full road geometries, the figure is used to show the spatial footprint of each regional network.

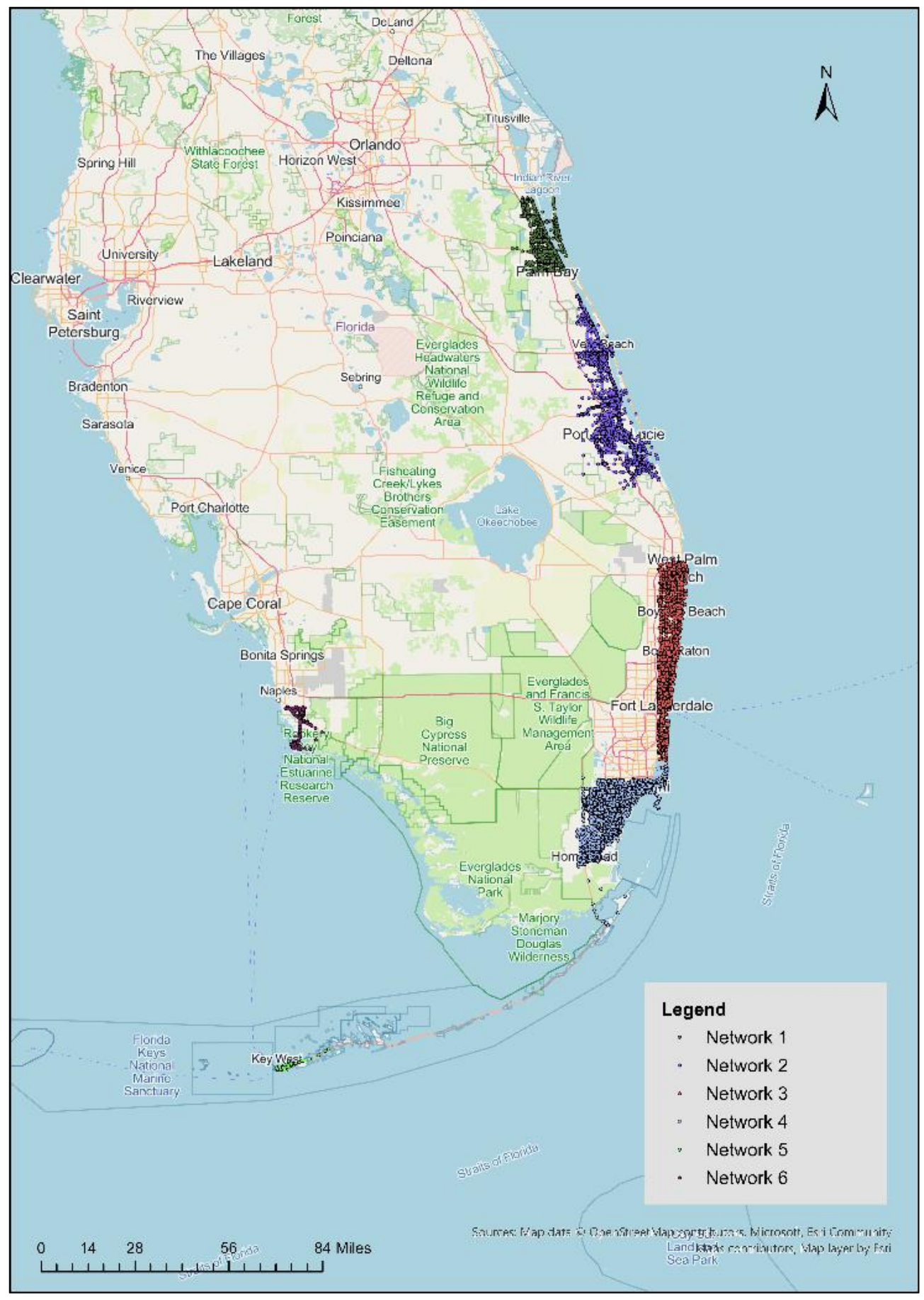

Figure 2. Spatial distribution of the six regional traffic networks in Florida.

*3.2 Network construction and preprocessing*

This subsection describes how the link travel time data are converted into the time-varying transport networks used by the methodological framework. For each region, the node set $V$ is formed from the origin and destination nodes in the data, and the directed link set $E$ is formed from the directed link records. The topology of each regional network is treated as fixed over the observation period, while the link operating conditions vary across observation times. Each available observation time $\tau$ therefore defines a network state $G(\tau)$ with the same topology and a different propagation strength matrix.

Preprocessing first identifies valid link-time observations. A valid observation has positive and finite link travel time and speed values. The invalid observation share is below 0.1% in all six data sets, as reported in Table 3. For each link, the scaling travel time $T_{ij}^{0}$ is estimated only from valid travel time observations. The small number of invalid link-time observations is replaced by the nearest valid observation on the same link when constructing the propagation strength matrix. This step preserves a complete propagation matrix at every available observation time.

The pre-processed link travel times are then normalised at the link level. For each directed link $(i,j)$. $T_{ij}^{0}$ is estimated from valid observations using Eq. (10), with $q=0.05$ in the main analysis. This link-specific scaling controls for differences in link length, road class, and normal operating speed. The normalised propagation strength $a_{ij}(\tau)$ is obtained from Eq. (11) for each available observation time.

After preprocessing, each regional data set is represented as a sequence of complete propagation strength matrices $\{A(\tau)\}$

over the available observation times. These matrices provide the empirical input for the DPC, and critical-area computations defined in Section 2.

*3.3 Spatial region design*

Critical-area identification requires a set of candidate spatial regions for each transport network. The main analysis divides the spatial extent of each regional network into regular grid cells based on representative link coordinates. Regular grids provide a transparent and reproducible spatial partition and allow consistent analysis across the six regional networks. The analysis retains grid cells that contain at least one representative link coordinate as candidate regions.

Each directed link is assigned to a candidate region according to its representative coordinate $z_{ij}$, following the link-to-region assignment in Eq. (13). The resulting set $E_{R_m}$ contains the links located in region $R_m$. Regional degradation is then applied to all links in $E_{R_m}$ when computing the regional DPC impact. This construction connects the spatial partition with the critical-area score while preserving the link-level traffic information in the original data.

The framework requires only a predefined set of candidate regions and a rule for assigning links to these regions. This empirical study uses regular grids because they can be constructed directly from the available link coordinates. When other spatial boundaries are available, such as traffic management zones, administrative districts, evacuation zones, or flood-risk zones, the same analysis workflow can use them as candidate regions.

Figure 3 illustrates the spatial region design using Network 4 as an example. The figure shows all non-empty grid cells to illustrate the spatial partition. Empty grid cells do not enter the critical-area scoring procedure. In the empirical scoring procedure, only grid cells with at least $n_{\min} = 10$ assigned links enter the main ranking. This minimal support requirement keeps the main ranking focused on candidate areas with enough observed links to represent a regional perturbation.

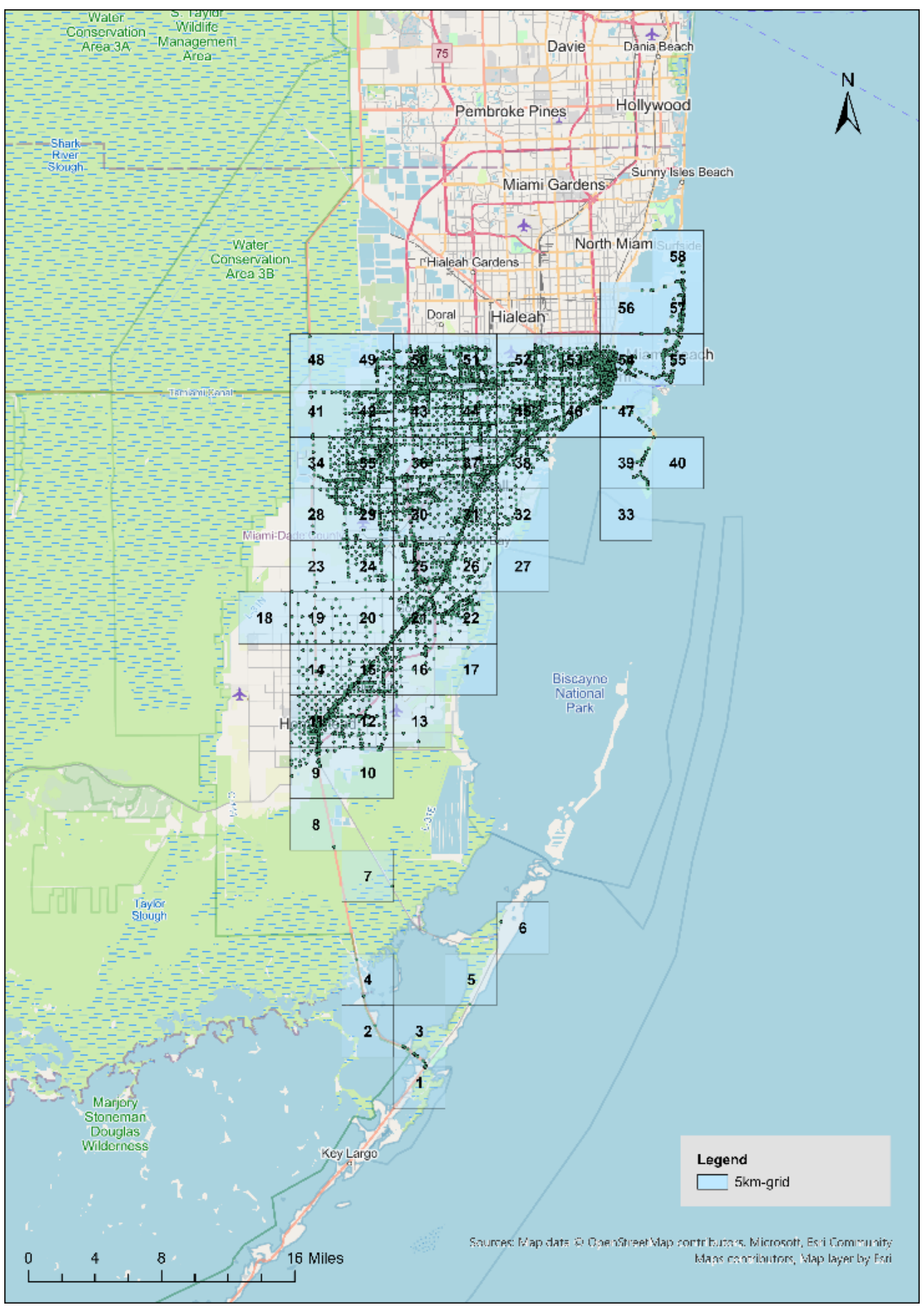

Figure 3. Spatial region design for critical-area identification.

Candidate region IDs follow the grid-cell order in each network map. The cells are numbered from south to north, and within each horizontal row from west to east.

*3.4 Experimental settings and comparison metrics*

The empirical evaluation starts with a benchmark validation experiment. The benchmark network contains a known structurally critical area, so it checks whether the regional DPC score can recover a known critical area before the analysis applies the method to the six Florida networks. The six regional networks then form the main large-scale empirical application.

Table 4 consolidates the experimental settings used in the benchmark validation and the Florida empirical analysis. The main specification uses $q = 0.05, \alpha = 0.5$ , 5 km by 5 km regular grid cells, and $n_{\min} = 10$ . These settings define the reference configuration for the main results, while the robustness analysis varies $q, \alpha$ and grid size.

Table 4. Experimental settings and comparison metrics.

| Component | Main setting | Role in the analysis |
|---|---|---|
| Benchmark validation | Benchmark network with a known critical area | Checks whether the regional DPC score can recover a known critical area |
| Empirical networks | Six Florida regional networks | Provides the main large-scale empirical application |
| Travel-time scaling | $q = 0.05$ | Estimates the link-specific scaling travel time $T_{ij}^0$ |
| Spatial partition | 5 km by 5 km regular grid cells | Defines candidate spatial regions |
| Minimum link support | $n_{\min}=10$ assigned links | Keeps ranked candidates at the area level |
| Regional degradation | $\alpha=0.5$ | Applies a moderate reduction to propagation strengths in a candidate region |
| Main outputs | $DPC(\tau), S_m(\tau,\alpha), \bar{S}_m(\alpha), S_m^{\max}(\alpha), F_m(\alpha)$ | Tracks global propagation dynamics and maps critical areas |
| Classical percolation baseline | Largest connected component loss after regional link removal | Measures topological disconnection under binary regional failure |
| Spectral topology baseline | Algebraic connectivity loss after regional link removal | Measures spectral connectivity change under regional removal |
| Conventional metrics | Link count, local speed degradation, aggregated edge betweenness | Provides size, local traffic, and shortest-path baselines |
| Robustness checks | $q,\alpha$ and grid size | Tests sensitivity of the main findings |
| Computational feasibility | Computation time by network | Assesses scalability on time-varying regional networks |

For each empirical network, the retained candidate regions receive a regional score at every available observation time. Since each network contains 1,281 available observation times, the analysis reports both time-varying results and summary measures. Let $\mathrm{T}$ denote the set of available observation times. The time-averaged regional score is:

$$\bar{S}_m(\alpha)=\frac{1}{|\mathrm{T}|}\sum_{\tau\in\mathrm{T}}S_m(\tau,\alpha) \tag{28}$$

The peak regional score is:

$$S_m^{\max}(\alpha)=\max_{\tau\in\mathrm{T}}S_m(\tau,\alpha) \tag{29}$$

Let $\mathrm{R}^+$ denote the retained candidate regions after the minimum link-support screening. The top-decile frequency records the share of observation times in which a candidate region belongs to the top 10 percent of retained regions by score:

$$F_m(\alpha)=\frac{1}{|\mathrm{T}|}\sum_{\tau\in\mathrm{T}}\mathbf{1}\left\{S_m(\tau,\alpha)\geq Q_{0.9}\left(\left\{S_r(\tau,\alpha):R_r\in\mathrm{R}^+\right\}\right)\right\} \tag{30}$$

where $\mathbf{1}\{\cdot\}$ is the indicator function. These quantities are temporal summaries of the same regional score $S_m(\tau,\alpha)$. The average score (Eq. 28) shows which areas remain important over the observation period, the peak score (Eq. 29) shows which areas have the largest short-term impact, and the top-decile frequency (Eq. 30) shows which areas often appear among the highest-scoring regions.

## 4. Empirical results

### *4.1 Benchmark validation on the Sioux Falls network*

Before applying the proposed method to the Florida networks, this subsection uses the standard Sioux Falls network as a benchmark validation case. The network contains 24 nodes and 76 directed links and has been widely used in traffic assignment and transport network studies (Meng et al., 2001; Meng and Yang, 2002). The purpose of this experiment is to check whether the regional DPC score can identify a structurally defined reference corridor in a small network whose topology can be inspected directly.

The benchmark uses the standard Sioux Falls topology and assigns unit propagation strength to all directed links. This setting focuses the experiment on the identification mechanism. The calculation uses the internal propagation time $t$ in the DPC definition, and DPC is obtained from the long-time average of the propagation state under the static operator. The observed time index $\tau$ enters the Florida application, where each five-minute traffic state provides $a_{ij}(\tau)$.

The benchmark experiment follows the same regional scoring logic as the empirical analysis. The schematic network is divided into candidate regions, and each candidate region is degraded by setting $\alpha = 0.5$. The reference corridor is fixed before calculating the scores and is denoted by $R_1$.

Figure 4 reports the benchmark validation result. Panel (a) shows the predefined reference corridor $R_1$, where the red links indicate the propagation couplings degraded in the validation experiment. Panel (b) shows the regional DPC score over the candidate regions. Panel (c) ranks all candidate regions by their regional DPC scores. Panel (d) reports the baseline DPC and the degraded DPC after applying the regional degradation to $R_1$.

The highest-scoring region is $R_1$. The baseline DPC is 14.084, and degrading $R_1$ reduces the DPC to 12.863. The relative DPC loss is 0.087. These results show that the regional DPC score identifies the reference corridor and quantifies its impact on global propagation coverage.

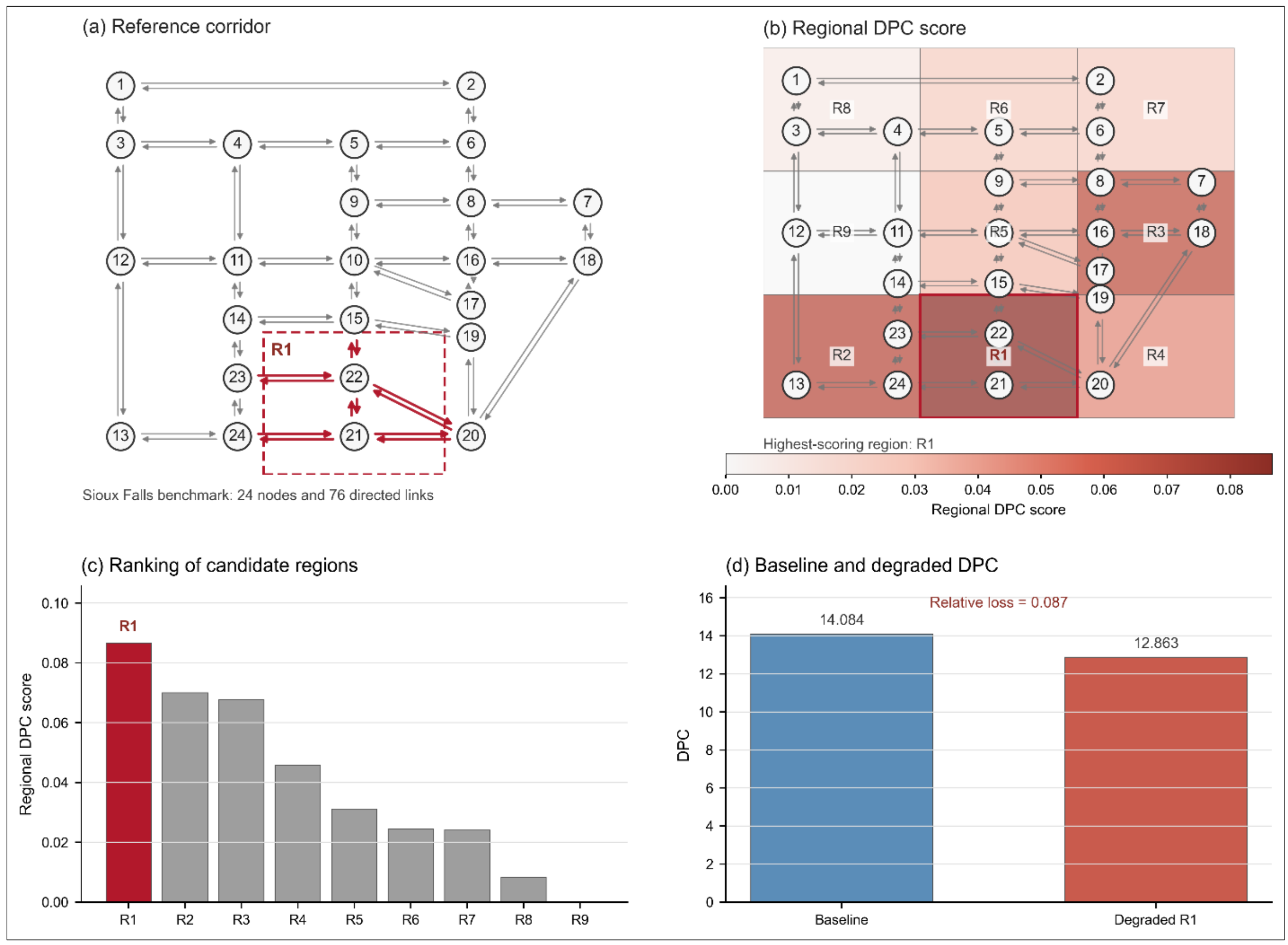

Figure 4. Benchmark validation on the Sioux Falls network.

The benchmark provides a preliminary validation of the critical-area identification procedure. The following subsections apply the same workflow to the six Florida networks and examine how critical areas evolve under observed time-varying travel conditions. The comparison with classical percolation and spectral-topological baselines is reported in Section

## *4.2 Global DPC evolution in the Florida networks*

This subsection examines the global DPC dynamics in the six Florida networks before mapping critical areas. For each available observation time, the link travel times are converted into propagation strengths, and the corresponding propagation operator gives $DPC(\tau)$. Since the six networks have different sizes, Figure 5 reports normalised DPC values.

The normalisation divides each DPC value by the maximum DPC observed in the same network. Thin blue lines report the computed values at all 1,281 available observation times, and red lines show smoothed trends. The curves confirm that the observed link travel time data produce clear temporal variation in propagation connectivity. Networks 2, 3, and 4 remain relatively stable, with minimum normalised DPC values between 0.922 and 0.929. Networks 5 and 6 show larger relative fluctuations, with minimum values of 0.846 and 0.839, respectively.

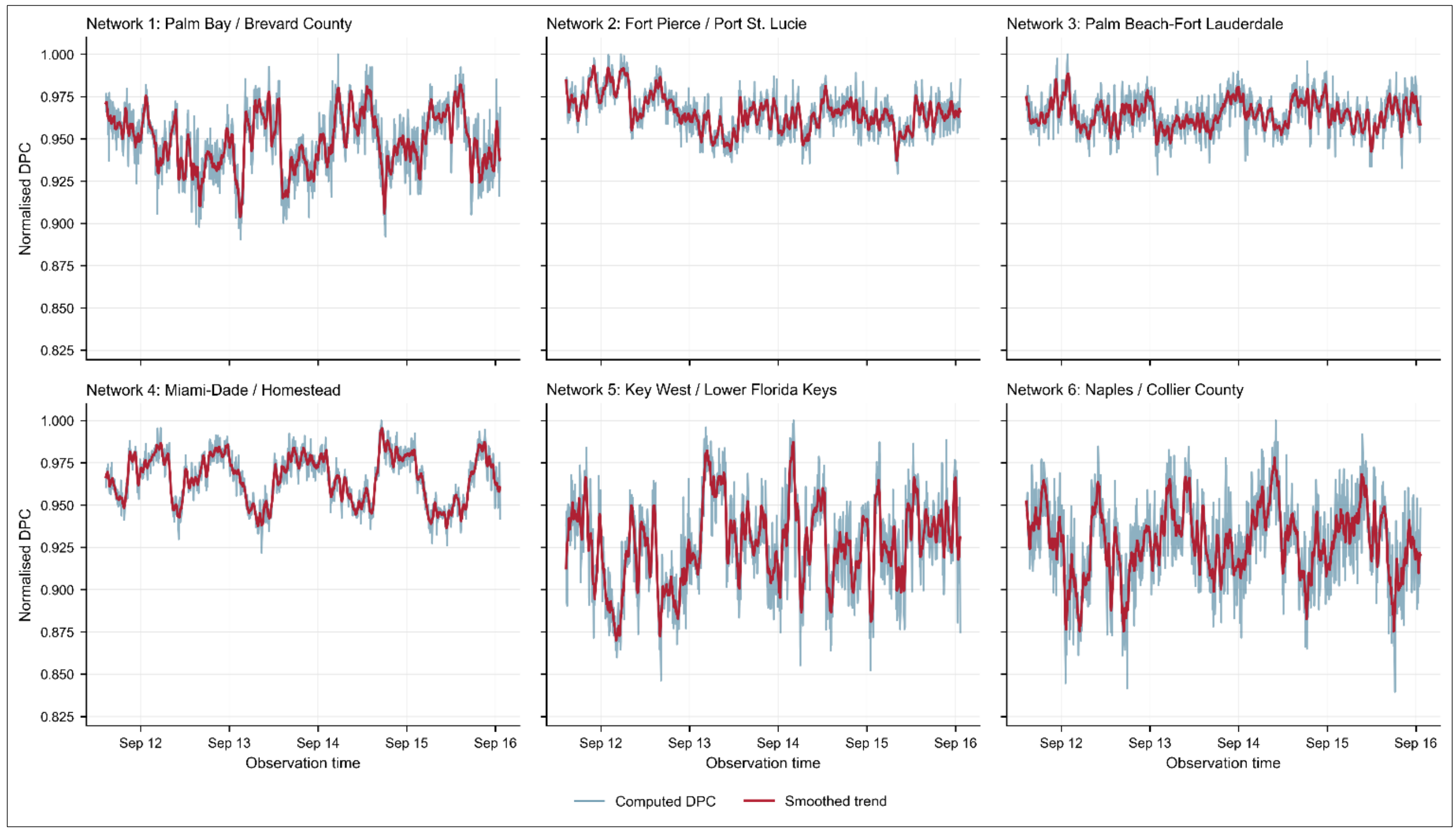

Figure 5. Time-varying DPC in the six Florida networks.

Table 5 summarises the temporal DPC patterns. Network 6 records the largest maximum decline, at 16.05 percent, followed by Network 5 at 15.38 percent and Network 1 at 10.96 percent. Networks 2, 3, and 4 record smaller maximum declines, all below 8 percent. The coefficient of variation complements the maximum-decline measure by summarising the overall relative fluctuation of each normalised DPC series. Networks 5 and 6 also have the largest coefficients of variation, at 0.0296 and 0.0271, which is consistent with the stronger short-term fluctuations visible in Figure 5. These differences indicate that the six regional networks experience different degrees of propagation-connectivity fluctuation under the observed traffic conditions.

Table 5. Temporal summary of normalised DPC in the six Florida networks.

| Network | Region | Mean normalised DPC | Minimum normalised DPC | Maximum decline (%) | Time of minimum DPC | Coefficient of variation |
|---|---|---|---|---|---|---|
| Network 1 | Palm Bay / Brevard County | 0.948 | 0.890 | 10.96 | 2017-09-13 03:05 | 0.0197 |
| Network 2 | Fort Pierce / Port St. Lucie | 0.966 | 0.929 | 7.07 | 2017-09-15 08:20 | 0.0123 |
| Network 3 | Palm Beach-Fort Lauderdale corridor | 0.964 | 0.929 | 7.12 | 2017-09-13 02:05 | 0.0108 |
| Network 4 | Miami-Dade / Homestead | 0.966 | 0.922 | 7.82 | 2017-09-13 08:45 | 0.0153 |
| Network 5 | Key West / Lower Florida Keys | 0.927 | 0.846 | 15.38 | 2017-09-12 16:15 | 0.0296 |
| Network 6 | Naples / Collier County | 0.928 | 0.839 | 16.05 | 2017-09-15 18:20 | 0.0271 |

The global DPC results show when each regional network experiences lower propagation connectivity and how large the decline is relative to its own highest observed DPC.

*4.3 Critical-area identification in the Florida networks*

This subsection applies the regional DPC score to the six Florida networks. The calculation uses all 1,281 available observation times recorded at five-minute intervals. For each candidate region $R_m$, the propagation strengths of links inside the region are reduced by $\alpha = 0.5$, and the DPC loss is computed from the degraded propagation operator. The regional scores are then summarised by the time-averaged score $\bar{S}_m$, the peak score $S_m^{\max}$, and the top-decile frequency $F_m$.

Figure 6 shows that the regional DPC score identifies high-score areas in all six networks. The highest-ranked regions indicate locations where a regional reduction in propagation strength produces large losses of global DPC. This result gives a direct spatial diagnosis of transport network vulnerability, because the score evaluates the network-wide consequence of a regional degradation event.

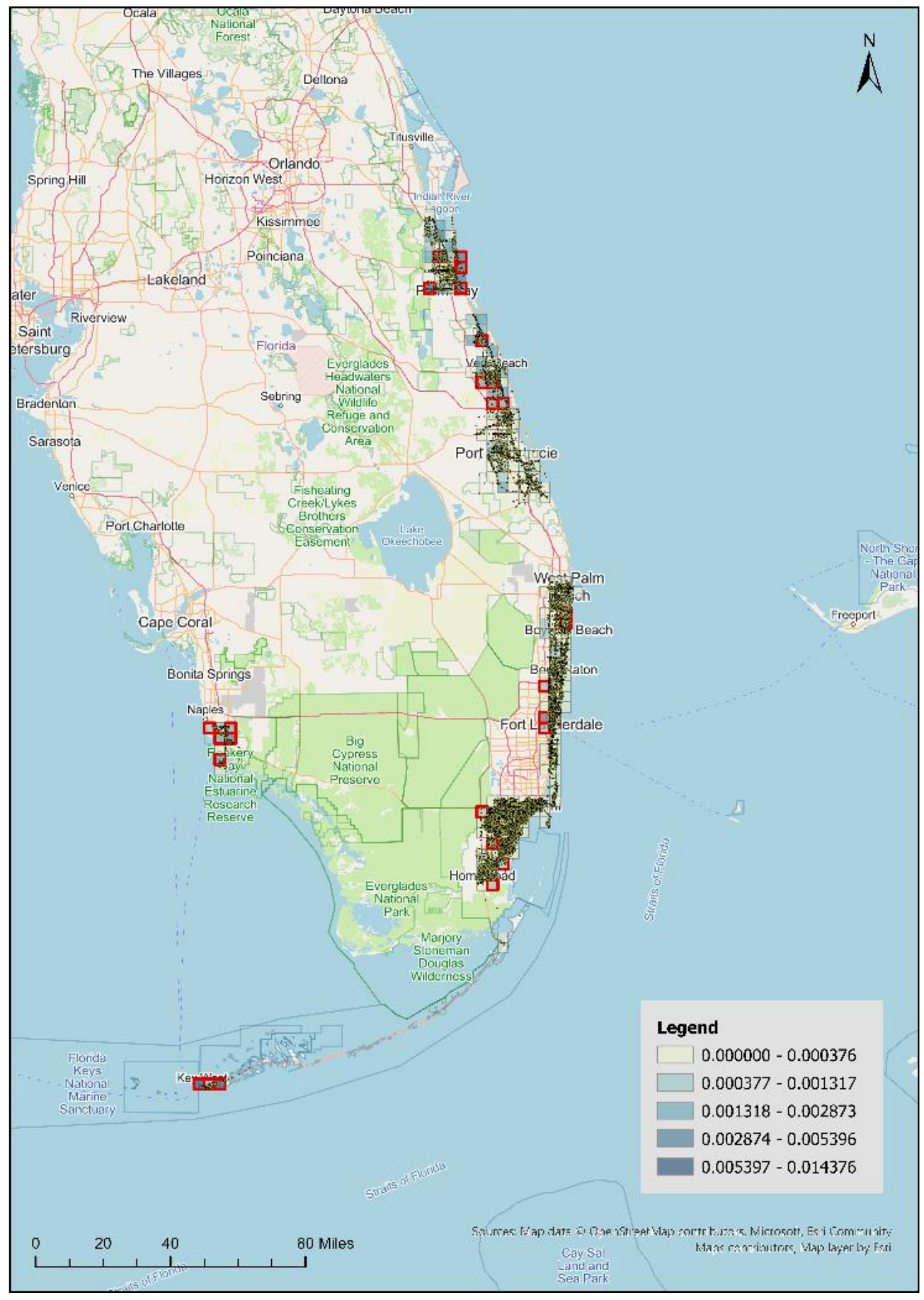

Figure 6. Statewide overview of critical areas identified across the six Florida networks. (Grid cells are coloured by the time-averaged regional DPC score $\bar{S}_m$ , Darker cells indicate larger DPC losses after regional degradation. Points denote link-level representative coordinates. Red outlines indicate the top-five candidate regions ranked by $\bar{S}_m$ within each network. Detailed network-specific maps are provided in Appendix A.

Table 6 reports the highest-ranked candidate region in each network. Network 3 has the clearest dominant critical area. Area 12 records a mean score of 0.00326 and appears in the top decile in 73.8 percent of the observation times. Its mean score is 3.77 times that of the second-ranked region. Network 2 and Network 6 also show clear leading regions, with top-to-second score ratios of 1.52 and 1.98, respectively.

Table 6. Highest-ranked critical areas identified by the regional DPC score.

| Network | Region | Top area | Links in top area | $\bar{S}_m$ | $S_m^{\max}$ | $F_m$ | Top-to-second ratio |
|---|---|---|---|---|---|---|---|
| Network 1 | Palm Bay / Brevard County | Area 13 | 76 | 0.01113 | 0.04838 | 0.614 | 1.10 |
| Network 2 | Fort Pierce / Port St. Lucie | Area 51 | 55 | 0.00436 | 0.01338 | 0.625 | 1.52 |
| Network 3 | Palm Beach-Fort Lauderdale corridor | Area 12 | 51 | 0.00326 | 0.01532 | 0.738 | 3.77 |
| Network 4 | Miami-Dade / Homestead | Area 9 | 59 | 0.00100 | 0.00558 | 0.437 | 1.01 |
| Network 5 | Key West / Lower Florida Keys | Area 1 | 31 | 0.01438 | 0.07956 | 0.637 | 1.14 |
| Network 6 | Naples / Collier County | Area 5 | 61 | 0.00540 | 0.06154 | 0.255 | 1.98 |

Note: The top-to-second ratio is computed from the time-averaged regional DPC score. For networks with fewer than ten retained candidate regions, $F_m$ records the share of observation times in which the region has the highest score among retained candidate regions.

Network 1 and Network 5 also identify stable high-impact regions, although their leading regions are close to the second-ranked regions. In Network 1, Area 13 has a mean score of 0.01113, while Area 1 has a mean score of 0.01009. In Network 5, Area 1 and Area 3 have mean scores of 0.01438 and 0.01260. These cases show that the leading area remains identifiable, while other high-ranking regions also make substantial contributions to propagation loss.

Network 4 shows a more distributed pattern. Area 9 ranks first, with a mean score of 0.00100, while Area 14 follows closely with a mean score of 0.00099. The top-to-second ratio is 1.01, which indicates that the Miami-Dade and Homestead network contains several comparable high-impact regions. This pattern is consistent with a dense urban network where multiple nearby areas can produce similar losses in DPC.

The results also show that the regional DPC score captures more than local link density. In Network 3, several dense coastal grid cells do not enter the top-five set, while Area 12 produces the largest average DPC loss. Area 23 in Network 4 contains 470 links and ranks third, while Area 9 contains 59 links and ranks first. In Network 5, the largest candidate region contains 88 links and ranks below two smaller regions. The regional DPC score therefore identifies areas with large global propagation consequences, even when those areas are not the largest local clusters of links.

Overall, Figure 6 and Table 6 show that the proposed regional DPC score provides a spatial diagnosis of transport network vulnerability. The method identifies where regional degradation causes the largest loss of DPC, and it reveals different spatial patterns across the six networks.

*4.4 Comparison with classical percolation and conventional indicators*

This subsection compares the regional DPC score with classical percolation and conventional indicators. Classical percolation evaluates a binary connectivity consequence: a set of links is removed and the largest connected component records how much of the network remains connected (Albert et al., 2000; Cohen et al., 2000). The same regional-removal logic also appears in area-covering road network vulnerability analysis (Jenelius and Mattsson, 2012). The regional DPC score evaluates a different consequence. It weakens the propagation strengths inside a candidate region and measures the resulting loss of effective propagation coverage.

For each retained candidate region, the classical percolation baseline removes the region links from the underlying undirected network and computes the relative loss of the largest connected component. The spectral-topological baseline computes the relative loss of algebraic connectivity after the same removal operation. The additional local baselines are link count, the mean local degradation $1-a_{ij}(\tau)$, and aggregated directed edge betweenness (Girvan and Newman, 2002).

All rankings use larger values as more critical. Figure 7 reports ranking agreement with DPC using Spearman correlation and top-decile overlap. Table 7 reports the top-ranked area selected by each method.

Figure 7 shows that classical percolation and regional DPC often rank candidate regions differently. In Networks 1 to 4, the Spearman correlations between DPC and classical percolation are negative, ranging from -0.52 to -0.23, and the top-decile overlap rates are 0.00, 0.17, 0.00 and 0.20, respectively.

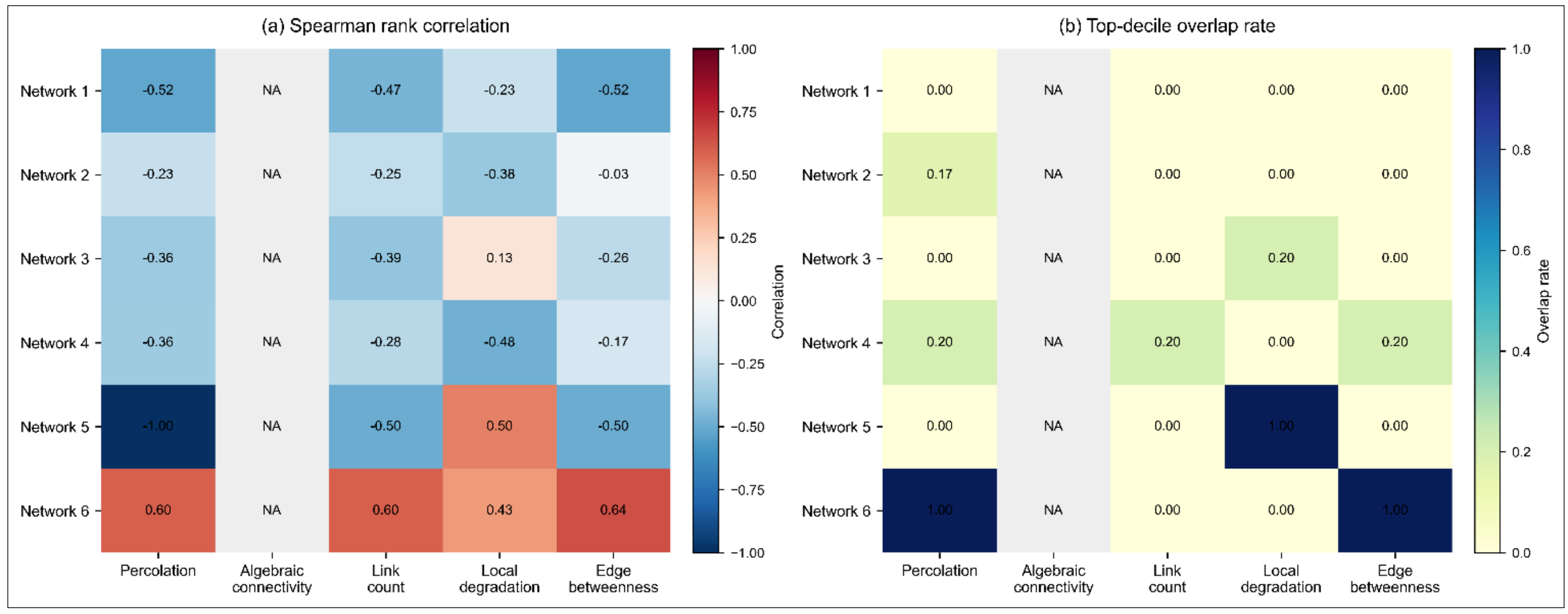

Figure 7. Agreement between regional DPC and baseline rankings. (a) Spearman rank correlation between the DPC ranking and each baseline ranking. (b) Top-decile overlap rate between the highest-ranked DPC regions and the highest-ranked baseline regions. Grey cells labelled NA indicate a baseline with no effective ranking because all candidate regions are tied.

Table 7 gives the same pattern at the top-area position. Percolation selects Area 10, Area 13 and Area 40 in Networks 1, 3 and 4, while DPC selects Area 13, Area 12 and Area 9. Network 2 is a partial exception because both methods select Area 51 as the top area, although the rank correlation remains weak. Network 6 also shows agreement between DPC and classical percolation, which indicates that the two views can coincide in small or structurally constrained networks.

Table 7. Top-ranked candidate areas under DPC and baseline indicators.

| Network | Region | DPC | Classical percolation | Algebraic connectivity | Link count | Local degradation | Edge betweenness |
|---|---|---|---|---|---|---|---|
| Network 1 | Palm Bay / Brevard County | Area 13 | Area 10 | Tie (24) | Area 10 | Area 5 | Area 9 |
| Network 2 | Fort Pierce / Port St. Lucie | Area 51 | Area 51 | Tie (54) | Area 11 | Area 46 | Area 20 |
| Network 3 | Palm Beach-Fort Lauderdale corridor | Area 12 | Area 13 | Tie (48) | Area 26 | Area 2 | Area 26 |
| Network 4 | Miami-Dade / Homestead | Area 9 | Area 40 | Tie (43) | Area 23 | Area 41 | Area 25 |
| Network 5 | Key West / Lower Florida Keys | Area 1 | Area 2 | Tie (3) | Area 2 | Area 1 | Area 3 |
| Network 6 | Naples / Collier County | Area 5 | Area 5 | Tie (8) | Area 7 | Area 4 | Area 5 |

Note: Tie indicates that all retained candidate regions have the same algebraic-connectivity loss after regional link removal.

The spectral-topological baseline shows a different limitation. Regional link removal makes algebraic connectivity drop to zero for all retained candidate regions in each network. The resulting ties produce no meaningful ranking, shown as NA in Figure 7 and as all-region ties in Table 7. Link count, local degradation and edge betweenness also select different top areas in most networks. In Network 3, link count and edge betweenness select Area 26, while DPC selects Area 12. In Network 4, link count selects Area 23, local degradation selects Area 41 and edge betweenness selects Area 25, while DPC selects Area 9.

These results clarify the methodological difference between classical percolation and quantum-percolation-based DPC. Classical percolation identifies regions whose removal fragments binary connectivity. Regional DPC identifies regions whose functional weakening causes the largest loss of effective propagation coverage. The two approaches answer different vulnerability questions. DPC extends the percolation logic from binary connectivity failure to continuous propagation degradation and provides spatial information beyond common topological, local traffic and shortest-path indicators.

*4.5 Robustness of critical-area identification*

This subsection examines whether the identified critical areas remain stable under alternative parameter settings. The main setting uses $q = 0.05, \alpha = 0.5$ , and 5 km by 5 km grid cells. The robustness analysis changes one element at a time. It uses $q = 0.03$ and $q = 1.0$ to test travel-time scaling, $\alpha = 0.3$ and $\alpha = 0.7$ to test the regional degradation strength, and 4 km and 6 km grid cells to test the spatial partition. All tests use the same regional DPC scoring procedure and the same available observation times.

For settings that keep the 5 km grid, the candidate regions have matching Area IDs, so rank agreement is measured by the Spearman rank correlation between the alternative ranking and the main ranking. For all settings, including alternative grid sizes, spatial stability is measured by the top-decile spatial overlap with the main setting. This second measure compares where the highest-scoring regions are located, which remains meaningful when the grid cells change. Figure 8 shows strong robustness across the six networks. In Panel (a), all Spearman correlations are at least 0.76, and most values exceed 0.90. The strongest stability appears in Network 3, where the rank agreement ranges from 0.94 to 0.98 across the four same-grid parameter settings. Networks 1, 2, 5, and 6 also show high agreement, with minimum correlations between 0.84 and 0.92. Network 4 has the lowest rank agreement, 0.76 under $\alpha = 0.3$, which is consistent with the more distributed critical-area pattern reported in Section 4.3.

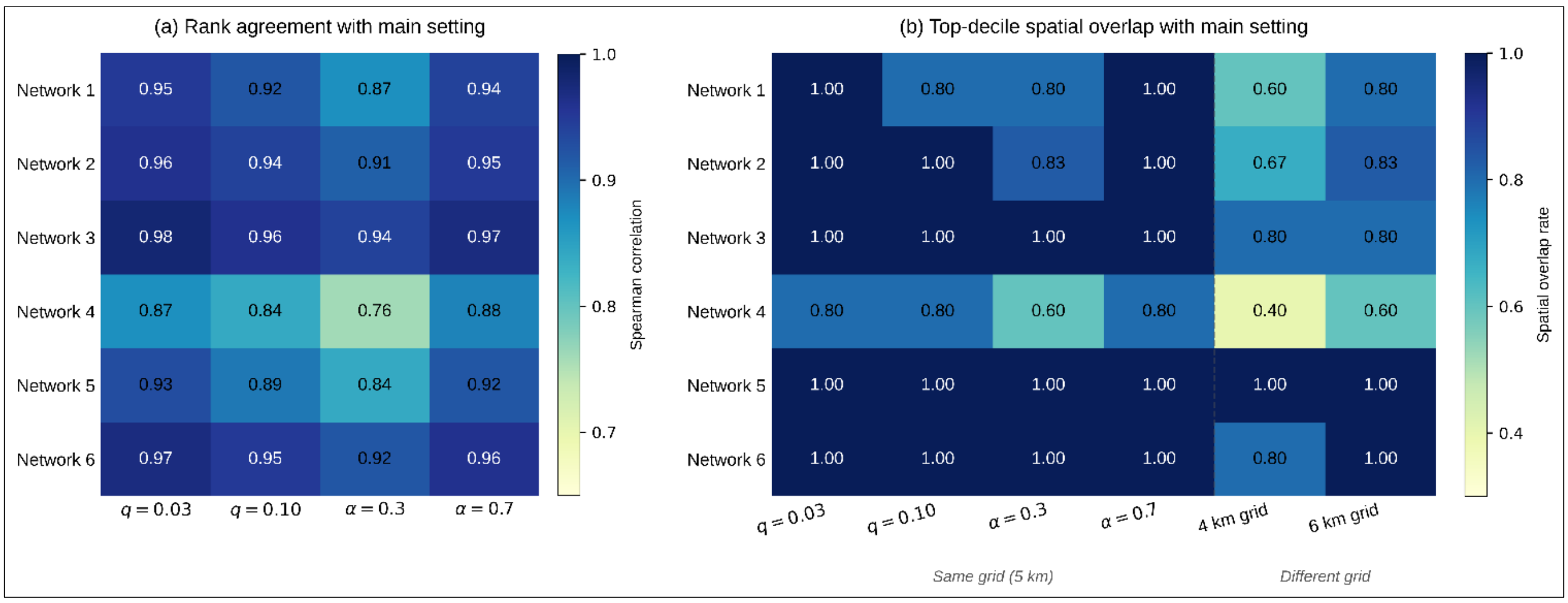

Figure 8. Robustness of critical-area identification under alternative parameter settings (Each robustness test changes only the parameter named on the x-axis. The 4 km and 6 km grid tests keep $q = 0.05, \alpha = 0.5$ ).

Panel (b) shows that the high-score areas also remain spatially stable. The top-decile spatial overlap is at least 0.80 for most network-setting combinations. Network 5 has perfect overlap under all alternative settings, and Networks 3 and 6 retain mean spatial overlap values of 0.93 and 0.97. The weakest spatial overlap occurs in Network 4 under the 4 km grid, with a value of 0.40. This result indicates that the finer spatial partition redistributes high scores among nearby urban cells in Miami-Dade and Homestead, while the broader high-impact zone remains visible.

Table 8 summarises the robustness patterns. The mean rank agreement ranges from 0.84 to 0.96, and the mean spatial overlap ranges from 0.67 to 1.00. These values show that the critical-area results have limited sensitivity to the travel-time scaling quantile, the regional degradation strength, and the grid size. The main critical-area patterns remain stable across the tested settings, with Network 4 showing the strongest spatial-scale sensitivity.

Table 8. Robustness summary of critical-area identification.

| Network | Region | Minimum rank agreement | Mean rank agreement | Weakest rank setting | Minimum spatial overlap | Mean spatial overlap | Weakest spatial setting |
|---|---|---|---|---|---|---|---|
| Network 1 | Palm Bay / Brevard County | 0.87 | 0.92 | $\alpha = 0.3$ | 0.60 | 0.83 | 4 km grid |
| Network 2 | Fort Pierce / Port St. Lucie | 0.91 | 0.94 | $\alpha = 0.3$ | 0.67 | 0.89 | 4 km grid |
| Network 3 | Palm Beach-Fort Lauderdale corridor | 0.94 | 0.96 | $\alpha = 0.3$ | 0.80 | 0.93 | 4 km and 6 km grids |
| Network 4 | Miami-Dade / Homestead | 0.76 | 0.84 | $\alpha = 0.3$ | 0.40 | 0.67 | 4 km grid |

| Network 5 | Key West / Lower Florida Keys | 0.84 | 0.90 | $\alpha = 0.3$ | 1.00 | 1.00 | All settings |
|---|---|---|---|---|---|---|---|
| Network 6 | Naples / Collier County | 0.92 | 0.95 | $\alpha = 0.3$ | 0.80 | 0.97 | 4 km grid |

Note: Rank agreement is measured by Spearman rank correlation between the main setting and the four alternative settings that use the same 5 km grid. Spatial overlap is measured by the top-decile spatial overlap between the main setting and all six alternative settings.

The full critical-area identification experiment uses all 1,281 available observation times and evaluates 230,580 region-time degradation cases across the six Florida networks. The sequential implementation requires approximately 16 h on a standard desktop workstation. The computation separates naturally by network, observation time, and candidate region, so parallel implementation can reduce the runtime. This runtime supports the retrospective empirical analysis, while operational deployment would benefit from parallel computing, sparse spectral approximation, and incremental updating across consecutive traffic states.

## 5. Discussion

The empirical results show that regional DPC translates time-varying link operating conditions into a spatial diagnosis of transport network vulnerability. The highest-scoring regions are the places where a reduction in local propagation strength produces the largest loss of network-wide effective propagation coverage. This finding matters because criticality does not follow directly from local link density, local speed deterioration, or shortest-path centrality. In several Florida networks, dense coastal grid cells and regions with many links do not receive the highest DPC scores, while smaller candidate regions can generate larger losses in global propagation connectivity. The proposed score therefore captures a propagation consequence that common local and topological indicators do not express.

This regional perspective complements the critical-link tradition in transport network vulnerability analysis. Earlier studies examined importance, exposure, accessibility loss, capacity reduction, and robustness at the link or segment scale (Jenelius et al., 2006; Sohn, 2006; Taylor et al., 2006; Sullivan et al., 2010). Those measures are valuable when an analyst wants to rank individual facilities. Transport agencies, however, often organise emergency response, evacuation control, infrastructure inspection, and recovery planning through spatial units such as districts, traffic management zones, flood-risk zones, and evacuation areas. A regional score fits this decision setting more naturally because it evaluates the joint effect of weakening all links located inside a spatial unit. The method therefore turns criticality from a list of isolated links into a map of high-impact areas that can support resource allocation and operational planning.

The comparison with classical percolation clarifies the methodological contribution. Classical percolation provides a powerful language for network robustness, from the original percolation process of Broadbent and Hammersley (1957) to the attack tolerance of complex networks studied by Albert et al. (2000) and Cohen et al. (2000). Traffic applications extended this logic to dynamical congestion and bottleneck formation, including percolation transitions in city traffic and heterogeneous flow networks (Li et al., 2015; Zeng et al., 2019; Hamedmoghadam et al., 2021; Zhu et al., 2025). These studies focus on connectivity loss, critical bottlenecks, or phase transitions under a removal or thresholding logic. The present results show a different vulnerability question. Regional DPC keeps the network topology fixed and weakens propagation strengths continuously, then measures the loss of effective propagation coverage. The low agreement between DPC and classical percolation in several networks shows that binary fragmentation and continuous propagation degradation can identify different critical areas.

The quantum-percolation interpretation gives the framework its additional structure. In classical graph measures, a connected node is typically counted as reachable once a path exists. In the DPC formulation, the propagation state evolves over the Hermitian propagation operator, and the time-averaged participation index measures how broadly the propagation state occupies the network. This links the transport application to the broader literature on localisation, quantum walks, and quantum network transport (Evers and Mirlin, 2008; Mülken and Blumen, 2011; Biamonte et al., 2019; Xu et al., 2021). The practical value of this formulation is that it treats link travel times as continuous propagation strengths. It can therefore represent functional degradation even when the network remains connected in a classical percolation sense. The result also differs from spectral topology alone. Algebraic connectivity describes structural cohesion, while regional DPC measures how a structured regional weakening changes effective propagation coverage under the observed traffic state.

The time-varying empirical setting strengthens the transport resilience interpretation. Resilience studies have emphasised that transport performance changes during disruption and recovery, and that vulnerability depends on both network structure and operating conditions (Bell, 2000; Faturechi and Miller-Hooks, 2014; Mattsson and Jenelius, 2015; Ganin et al., 2017). The present analysis follows this view by using observed five-minute link travel time data to update the propagation operator over time. The average regional score identifies areas with sustained influence over the observation period, the peak score identifies areas with severe short-duration impact, and the top-decile frequency identifies areas that repeatedly enter the highest-impact set. Together, these summaries show how critical areas can be studied as time-varying spatial objects. This makes the framework suitable for post-event diagnosis, resilience planning, and future monitoring systems that combine traffic-state data with regional vulnerability maps.

## 6. Conclusion

This paper proposes a dynamic propagation connectivity method based on quantum percolation for critical-area identification in transport networks. The method converts time-varying link travel times into continuous propagation strengths, constructs a propagation operator at each observation time, and measures the impact of each candidate region through a regional degradation experiment. The focus is the loss of network-wide effective propagation coverage caused by the joint weakening of links inside a spatial region. This gives transport network vulnerability analysis a regional diagnostic tool that differs from critical-link ranking.

The analysis gives three main findings:

(1) The Sioux Falls benchmark confirms that the regional DPC score can recover a structurally critical corridor. The predefined reference region $R_1$ receives the highest score. Degrading this region with $\alpha = 0.5$ reduces global DPC from 14.084 to 12.863, with a relative DPC loss of 0.087. This result shows that the regional degradation score reflects the contribution of a candidate region to global propagation coverage.

(2) The six Florida networks show clear spatial variation in critical areas. Using all 1,281 five-minute observation times, the empirical analysis identifies the candidate region with the largest average DPC loss in each network. Network 3 has the clearest dominant critical area. Its highest-scoring region has a mean score 3.77 times that of the second-ranked region and appears in the top decile in 73.8 percent of the observation times. Networks 2 and 6 also show clear leading regions. Networks 1, 4, and 5 show several high-impact regions with comparable scores, which indicates that regional vulnerability can have different spatial organisation across networks.

(3) The comparison with classical percolation, algebraic connectivity, and local indicators shows that regional DPC provides distinct vulnerability information. In Networks 1 to 4, the Spearman correlations between the DPC ranking and the classical percolation ranking are negative, and the top-decile overlap rates are low. Algebraic connectivity produces ties across all candidate regions under the regional removal experiment and therefore gives no effective ranking. Link count, local degradation, and edge betweenness also select different top-ranked regions in most networks. These results show that binary connectivity fragmentation, spectral-topological connectivity, local traffic deterioration, and shortest-path centrality cannot fully replace a regional score defined by propagation coverage loss.

The robustness analysis further supports the stability of the main results. Under alternative values of $q,\alpha$ and grid size, the mean rank agreement across the six networks ranges from 0.84 to 0.96, and the mean spatial overlap ranges from 0.67 to 1.00. Network 4 is the most sensitive case, which is consistent with its more distributed critical-area pattern and the small gap between its highest-ranked and second-ranked regions. The full experiment uses all 1,281 observation times and evaluates 230,580 region-time degradation cases, which supports large-scale retrospective analysis of time-varying transport networks.

Future research can extend this work in three directions. First, this study uses regular grids as candidate regions. Future applications can use traffic management zones, administrative districts, evacuation zones, or flood-risk zones to align the analysis more closely with operational boundaries. Second, the current computation uses a sequential implementation and requires approximately 16 h for the full experiment. Future work can improve computational efficiency through parallel computing, GPU acceleration, and incremental updating across consecutive observation times. As quantum computing hardware and quantum-inspired linear algebra methods continue to develop, the propagation-operator structure in this paper also provides a possible direction for new computational implementations. Third, future studies can embed the regional DPC score into optimisation models for emergency resource allocation, regional reinforcement, traffic control, and post-disruption recovery prioritisation. This would move critical-area identification from a diagnostic tool towards a decision-support method for improving transport network resilience.

## Appendix A. Network-specific critical-area maps

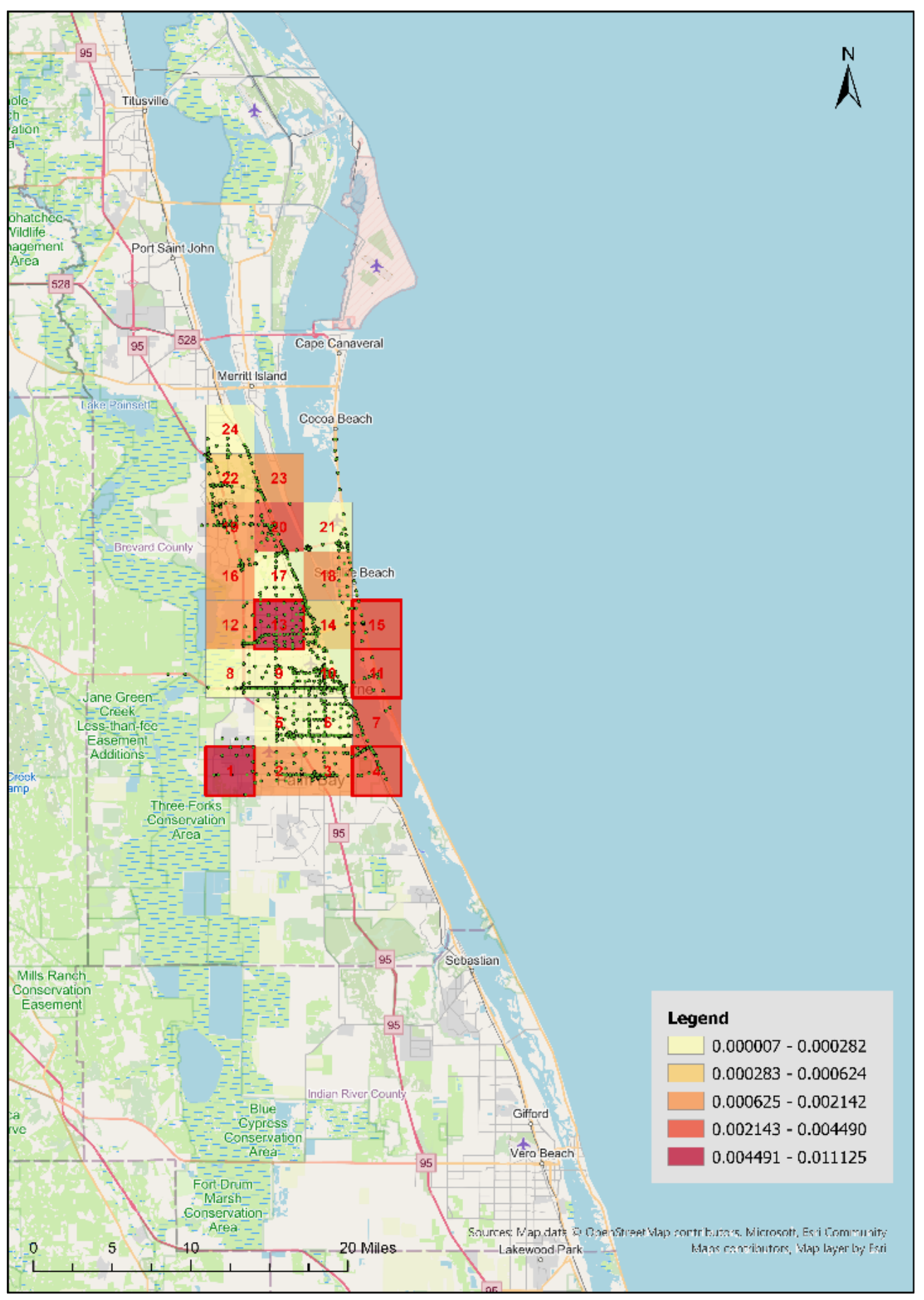

Figure A(a). Network 1: Palm Bay / Brevard County.

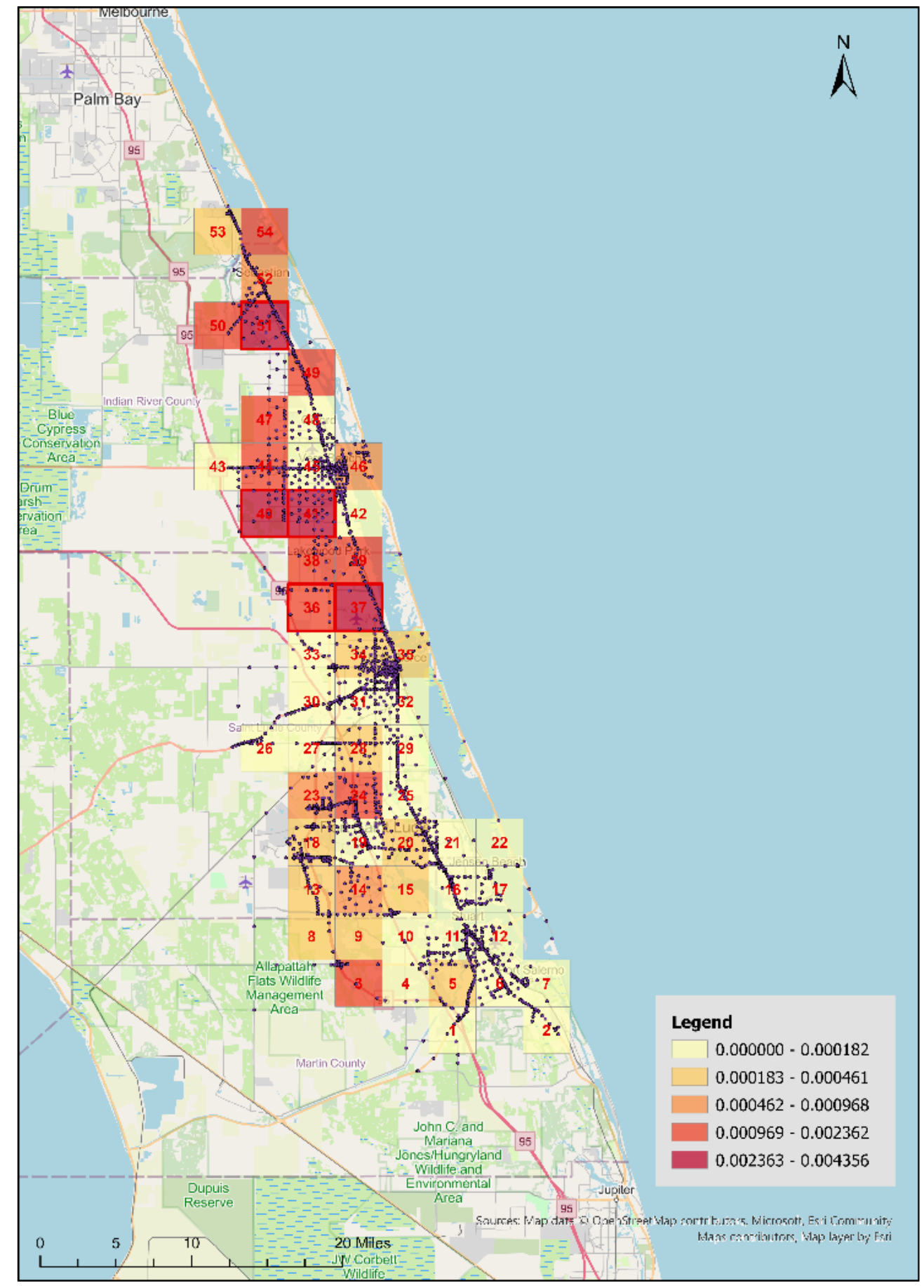

Figure A(b). Network 2: Fort Pierce / Port St. Lucie.

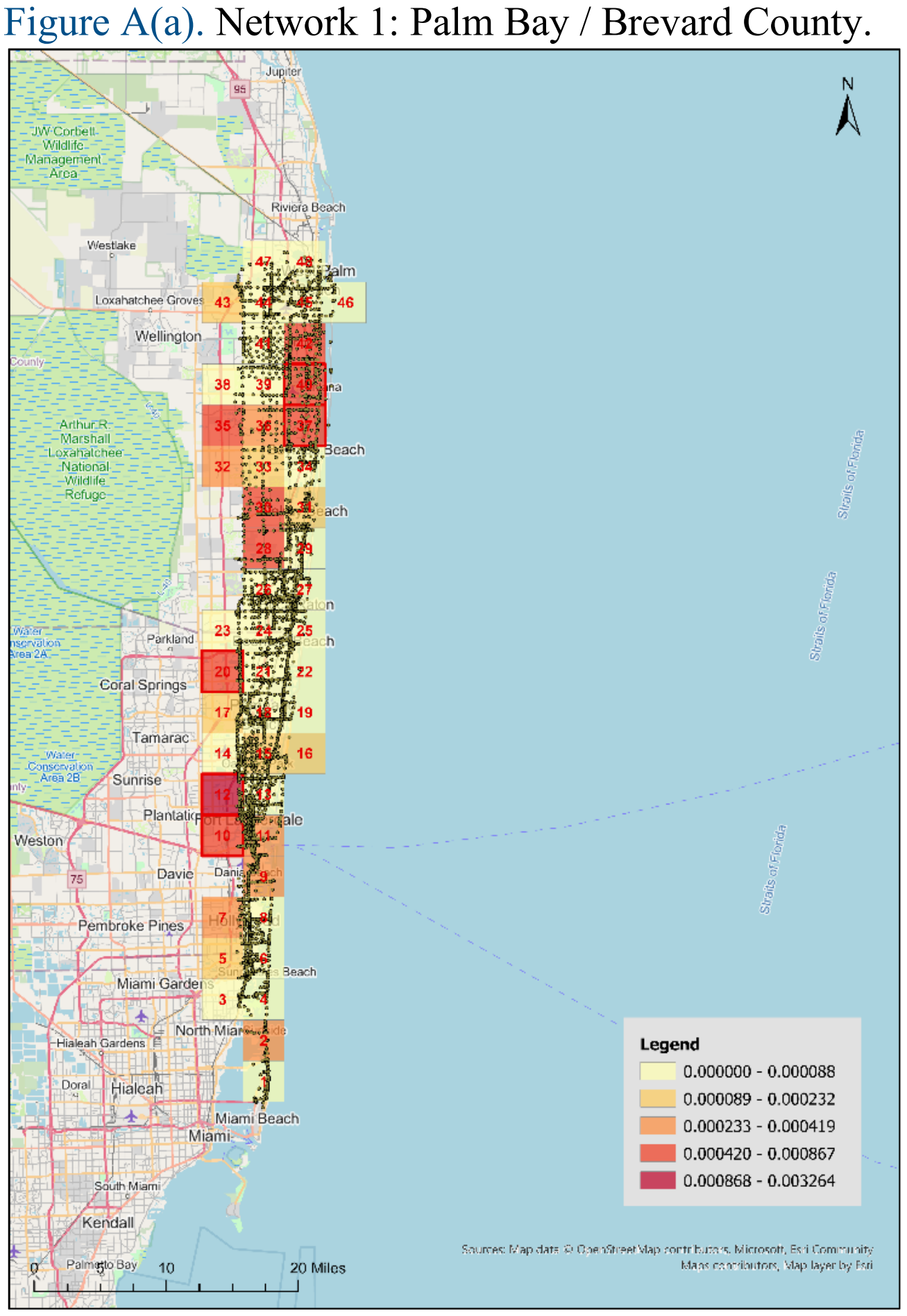

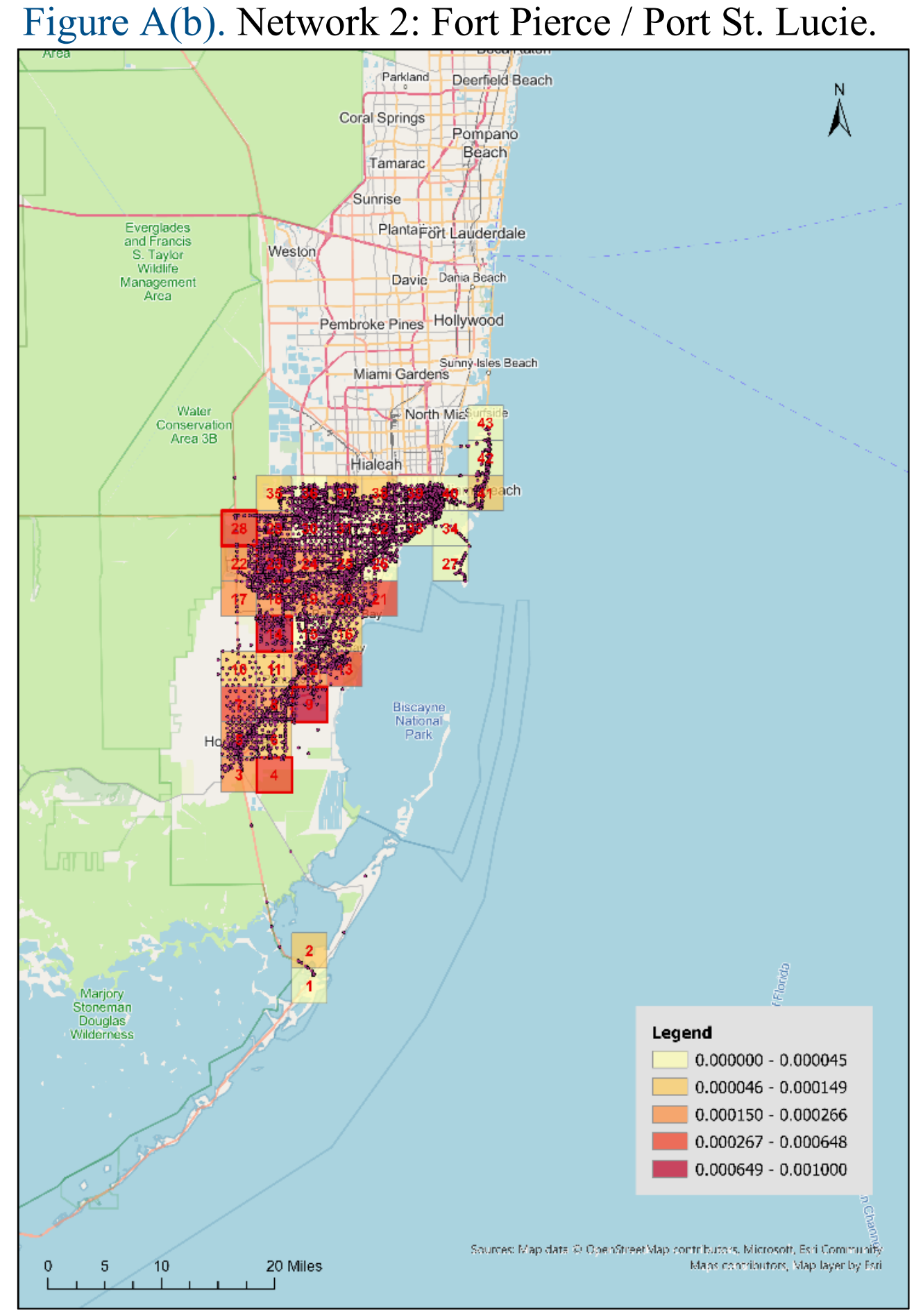

Figure A(c). Network 3: Palm Beach-Fort Lauderdale corridor. Figure A(d). Network 4: Miami-Dade / Homestead.

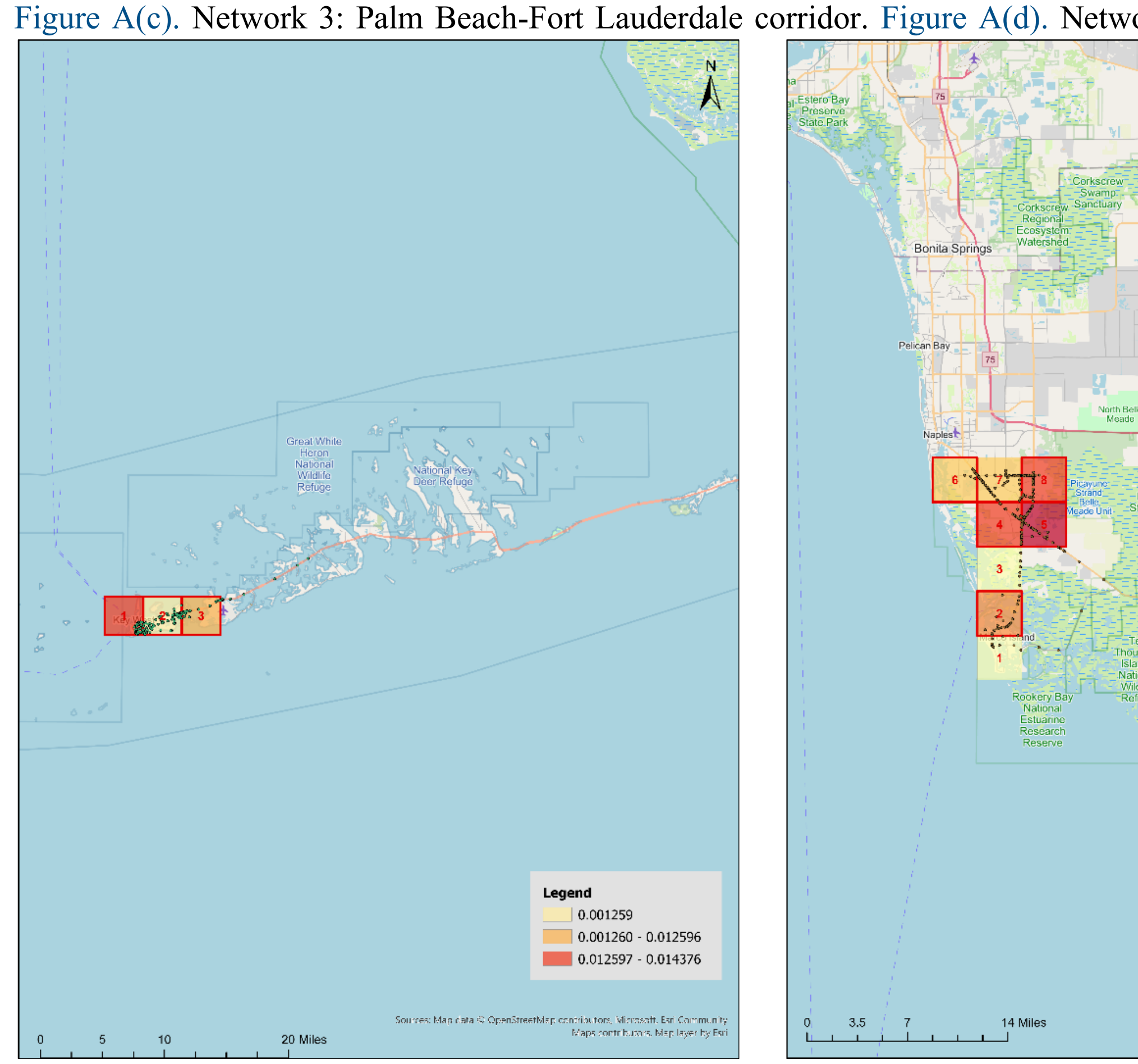

Figure A(e). Network 5: Key West / Lower Florida Keys. Figure A(f). Network 6: Naples / Collier County.

### Author contributions

**Junxiang Xu:** Writing – review & editing, Writing – original draft, Methodology, Formal analysis, Software, Conceptualization.

**Chence Niu:** Writing – review & editing, Methodology, Software.

**Tingting Zhang:** Writing – review & editing, Data processing.

**Divya Jayakumar Nair:** Review & editing, Supervision.

**Vinayak Dixit:** Review & editing, Supervision.

### Declaration of Interest Statement

The authors declare that they have no known competing financial interests or personal relationships that could have appeared to influence the work reported in this paper.

### Data Availability

Data will be made available on request.

### Declaration of generative AI and AI-assisted technologies in the writing process

During the preparation of this work, we employed ChatGPT-5.5 solely to assist with language editing and polishing. No content was generated by AI. After using this tool, we thoroughly reviewed and edited the content as necessary and accept full responsibility for the content of the published article.